\documentclass[aps,showpacs]{revtex4}

\usepackage{amsmath,amssymb,graphicx,float,color}

\begin{document}

\title{An empirically grounded agent based model for modeling directs, conflict detection and resolution operations in Air Traffic Management}

\author{C. Bongiorno$^{(1)}$, S. Miccich\`e$^{(1)}$, Rosario N. Mantegna$^{(1,2)}$}
\affiliation{$^{(1)}$Dipartimento di Fisica e Chimica, Universit\`a di Palermo, Viale delle Scienze, Ed. 18, I-90128, Palermo, Italy \\
                $^{(2)}$Center for Network Science and Department of Economics, Central European University, Nador 9, H-1051, Budapest, Hungary \\}

\date{\today}

%%%%%%%%%%%%%%%%%%%%%%%%%%%%%%%%%%%%%%%%%%%%%%%%%%%%%%%%%%%%%%%%%%%%%%%%%%%%%%%%%%%%%%%%%%%%%%%%%%%%
%%%%%%%%%%%%%%%%%%%%%%%%%%%%%%%%%%%%%%%%%%%%%%%%%%%%%%%%%%%%%%%%%%%%%%%%%%%%%%%%%%%%%%%%%%%%%%%%%%%%
\begin{abstract}

We present an agent based model of the Air Traffic Management socio-technical complex system that aims at modeling the interactions between aircrafts and air traffic controllers at a tactical level. The core of the model is given by the conflict detection and resolution module and by the directs module. Directs are flight shortcuts that are given by air controllers to speed up the passage of an aircraft within a certain airspace and therefore to facilitate airline operations. 
Conflicts resolution between flight trajectories can arise during the en-route phase of each flight due to both not detailed flight trajectory planning or unforeseen events that perturb the planned flight plan. Our model performs a local conflict detection and resolution procedure. Once a flight trajectory has been made conflict-free, the model searches for possible improvements of the system efficiency by issuing directs. 
We give an example  of model calibration based on real data. We  then provide an illustration of the capability of our model in generating scenario simulations able to give insights about the air traffic management system. We show that the calibrated model is able to reproduce the existence of 
a geographical localization of air traffic controllers' operations. Finally, we use the model to investigate  the relationship between directs and conflict resolutions (i) in the presence of perfect forecast ability of controllers, and (ii) in the presence of some degree of uncertainty in flight trajectory forecast. 
\end{abstract}

\maketitle

%%%%%%%%%%%%%%%%%%%%%%%%%%%%%%%%%%%%%%%%%%%%%%%%%%%%%%%%%%%%%%%%%%%%%%%%%%%%%%%%%%%%%%%%%%%%%%%%%%%%
%%%%%%%%%%%%%%%%%%%%%%%%%%%%%%%%%%%%%%%%%%%%%%%%%%%%%%%%%%%%%%%%%%%%%%%%%%%%%%%%%%%%%%%%%%%%%%%%%%%%
%%%%%%%%%%%%%%%%%%%%%%%%%%%%%%%%%%%%%%%%%%%%%%%%%%%%%%%%%%%%%%%%%%%%%%%%%%%%%%%%%%%%%%%%%%%%%%%%%%%%
\section{Introduction}
%%%%%%%%%%%%%%%%%%%%%%%%%%%%%%%%%%%%%%%%%%%%%%%%%%%%%%%%%%%%%%%%%%%%%%%%%%%%%%%%%%%%%%%%%%%%%%%%%%%%
%%%%%%%%%%%%%%%%%%%%%%%%%%%%%%%%%%%%%%%%%%%%%%%%%%%%%%%%%%%%%%%%%%%%%%%%%%%%%%%%%%%%%%%%%%%%%%%%%%%%
%%%%%%%%%%%%%%%%%%%%%%%%%%%%%%%%%%%%%%%%%%%%%%%%%%%%%%%%%%%%%%%%%%%%%%%%%%%%%%%%%%%%%%%%%%%%%%%%%%%%

In the future of air traffic management (ATM)  it is expected to observe an increase of traffic demand and new business challenges that will bring the current ATM system to its capacity limits. As a consequence, an overall productivity improvement is urgently needed \cite{sesar2007,comreg,eurocontrol2005,CWPP}. 
Within this major change not only the ATM productivity should be drastically enhanced, but consequently also the ATM system  safety and resilience standards will have to be improved. 

The way to improve resilience, safety and capacity in the future ATM domain 
goes through a better understanding of the actual air traffic system and its management procedures. 
Here we present an Agent Based Model (ABM) focused on aircrafts and controllers behavior acting in an area control center (ACC) of the ATM system. 
Our model helps in understanding the basis of the flight trajectory management and it can be used to perform scenario simulations aiming at investigating whether some modifications of the current operation rules can lead to an improvement in the general efficiency of the ATM system.

Agent based models started to become popular in the academy and research communities during the early nineties of the last century. During the nineties concepts and tools of complexity, chaos theory, computer science, and cellular automata were incorporated into the wave of development of Agent-Based Modeling \cite{heat}. Since these starting years the research field has expanded and evolved developing ABMs within several disciplines also focusing on topics as calibration and validation of the models \cite{windrum}. ABMs are a consolidated tool also in the Air Traffic Management domain. We can track down essentially three big research areas of application for the ABMs. In fact, we can have ABMs for the conflict detection and resolution \cite{kuchar}, for the management of the traffic flow \cite{agogino} and for the investigation of the aspects related to the role of human operators \cite{shah}. In this work we develop an ABM for conflict detection, conflict resolution and local enhancement of performances of specific flight trajectories.

ABMs for the conflict detection and resolution intervene at a tactical or pre-tactical level and provide methods for detecting and solving (multiple) conflicts on a pairwise or global basis. In Ref. \cite{kuchar} a set of categories has been proposed for the categorization of the different modeling approaches. For example, these categories include the dimensions at which the model works (vertical, horizontal or both), the method of conflict resolution (optimization, brute force, force field, \dots) and the type of maneuvers adopted by aircraft to avoid the conflict (vectoring, flight level changes and velocity changes). Among others, the work in Ref. \cite{durand} has received much attention, although it was not used at an operational level: it proposed a global method for conflict resolution based on genetic algorithms and taking into account future velocity uncertainties. The conflict detection and resolution algorithm of Ref. \cite{bilimoria1} is based on a local geometrical resolution of conflicts involving a combination of velocity changes and re-routings. Using the geometric characteristics of aircraft trajectories and intuitive reasoning, closed-form analytical solutions have been developed for optimal heading and/or speed commands. The conflict resolutions are optimal in the sense that they minimize the velocity vector changes required for conflict resolution, resulting in minimum deviations from the nominal trajectory. Another interesting approach is the conflict detection and resolution algorithm of Ref. \cite{eby} that is based on a potential field approach that in its original and simplistic version assumes that aircraft are like charged particles interacting in an external electric potential field. 

The agents of our ABM are aircraft/pilots and air traffic controllers who are active within an ACC in the European airspace. We simulate events that make a planned flight plan transform into an actual one. Therefore we are in the tactical phase of the air traffic management. The basic features of the model have been first introduced in Ref. \cite{sid2013}. A different implementation of the same basic principles has been proposed in Ref. \cite{sid2014monechi,plosmonechi}.

The model we discuss here is an evolution of the one of Ref. \cite{sid2013} and provides a conflict detection module based on the computation of the pairwise distances between all aircraft present in a certain airspace and also provides a module for the pairwise local resolution of conflicts at a tactical level. The approach of Ref. \cite{bilimoria1} is very close in spirit to the one presented here, where, however, the resolution algorithm works on the basis of numerical simulations. Our model performs a local conflict detection and resolution mainly based on geometric considerations and works at the level of an ACC. 

However, the controllers activity is not limited to the solution of possible safety events. In fact, one of the main tasks of each controller is that of facilitating the airlines operations. We have therefore implemented a module for the issuing of directs. Directs are flight trajectory modifications that are issued by controllers at a tactical level in order to speed up the passage of aircraft within a certain airspace and therefore to facilitate the airline operations. We believe that this is one of the novelties of our model, not present in those we have recalled above. Once trajectories have been made conflict-free, the controller searches for possible improvements of the system efficiency by issuing directs. The strategies by which directs are issued is the element that most characterizes the controllers' behavior. Indeed, the way we implement the strategies adopted by controllers to issue directs is based on information relative to each specific sector in the airspace as well as information relative to the entire considered airspace. 

Our model can be used to perform scenario simulations able to give insights about the ATM system. 
Specifically, we show what is the relationship between directs and conflict resolution events conditioned to assuming perfect or imperfect forecast ability of controllers.  This last issue might be relevant for understanding the evolution of the ATM system from the current to the so-called SESAR scenario \cite{sesar2007}. Indeed, a version of our model able to perform simulations relevant for the SESAR scenario has been provided in \cite{sid2015, pap49}.

The paper is organized as follows. In section \ref{datasec} we present the data used for the empirical characterization of the system and for model calibration. In section \ref{model} we describe the main features of the modules of the ABM. In section \ref{calibration} we summarize the model calibration. In section \ref{results} we show a few results obtained with our ABM using the model parameters obtained with the calibration procedure and in section \ref{directsafety} we show a few results obtained with our ABM and trying to exploit the possible range of model parameters. In section \ref{concl} we finally draw our conclusions.

%%%%%%%%%%%%%%%%%%%%%%%%%%%%%%%%%%%%%%%%%%%%%%%%%%%%%%%%%%%%%%%%%%%%%%%%%%%%%%%%%%%%%%%%%%%%%%%%%%%%
%%%%%%%%%%%%%%%%%%%%%%%%%%%%%%%%%%%%%%%%%%%%%%%%%%%%%%%%%%%%%%%%%%%%%%%%%%%%%%%%%%%%%%%%%%%%%%%%%%%%
%%%%%%%%%%%%%%%%%%%%%%%%%%%%%%%%%%%%%%%%%%%%%%%%%%%%%%%%%%%%%%%%%%%%%%%%%%%%%%%%%%%%%%%%%%%%%%%%%%%%
\section{Data} \label{datasec}
%%%%%%%%%%%%%%%%%%%%%%%%%%%%%%%%%%%%%%%%%%%%%%%%%%%%%%%%%%%%%%%%%%%%%%%%%%%%%%%%%%%%%%%%%%%%%%%%%%%%
%%%%%%%%%%%%%%%%%%%%%%%%%%%%%%%%%%%%%%%%%%%%%%%%%%%%%%%%%%%%%%%%%%%%%%%%%%%%%%%%%%%%%%%%%%%%%%%%%%%%
%%%%%%%%%%%%%%%%%%%%%%%%%%%%%%%%%%%%%%%%%%%%%%%%%%%%%%%%%%%%%%%%%%%%%%%%%%%%%%%%%%%%%%%%%%%%%%%%%%%%

The database we use combines together information obtained from DDR (Demand Data Repository)  \cite{DDR}  and NEVAC (Network Estimation Visualization of ACC Capacity) \cite{NEVAC} files already described in Ref. \cite{sid2011}. Data are collected by EUROCONTROL (http://www.eurocontrol.int), the European public institution that coordinates and plans air traffic control for all Europe \footnote{Data  were obtained as part of the {\em{SESAR Joint Undertaking}} WP-E  research project ELSA ``Empirically grounded agent based model for the future ATM scenario''. Data can be accessed by asking permission to the owner (EUROCONTROL).}.

The database includes all flights occurring in the enlarged European Civil Aviation Conference (ECAC) airspace {\footnote{Countries in the enlarged ECAC space are: Iceland (BI), Kosovo (BK), Belgium (EB), Germany-civil (ED), Estonia (EE), Finland (EF), UK (EG), Netherlands (EH), Ireland (EI), Denmark (EK), Luxembourg (EL), Norway (EN), Poland (EP), Sweden (ES), Germany-military (ET), Latvia (EV), Lithuania (EY), Albania (LA), Bulgaria (LB), Cyprus (LC), Croatia (LD), Spain (LE), France (LF), Greece (LG), Hungary (LH), Italy (LI), Slovenia (LJ), Czech Republic (LK), Malta (LM), Monaco (LN), Austria (LO), Portugal (LP), Bosnia-Herzegovina (LQ), Romania (LR), Switzerland (LS), Turkey (LT), Moldova (LU), Macedonia (LW), Gibraltar (LX), Serbia-Montenegro (LY), Slovakia (LZ), Armenia (UD), Georgia (UG), Ukraine (UK).}} even if they departed and/or landed in airports external to the enlarged ECAC airspace. In the present paper, we analyze data referring to the Aeronautical Information Regulation and Control (AIRAC)  cycle beginning on May 6th, 2010 and ending on June 3rd, 2010, corresponding to the AIRAC cycle number 334. The DDR data consists of the trajectories of flights, along with some additional information about flights (IATA code, type of aircraft, among the others). We use this additional information in order to set some filters. In fact, from the database we remove a certain number of flights that are not directly related to commercial air traffic management. For example we remove military flights in operation. Hereafter we list the filters we used to select our set of data:
(a) all flights occurring during weekdays; (b) flights having at least two points in the Italian ACC airspace (LIRR); (c) flights with at least two navigation points crossed at an altitude higher then 240FL in the flight plan; (d) only non-military and commercial flights having a IATA code; (e) only land-plane aircraft, i.e. no helicopter, gyrocopter, etc; (f) flights with a duration longer than 10 minutes. 

A planned or a realized trajectory is made by  a sequence of navigation points planned or crossed by the aircraft, together with altitudes and timestamps.  For each flight, we have access to two different flight plans: the last filled plan (labeled as M1 type file)  -- filed from 6 months to one or two hours before the real departure -- and the realized flight plan (labeled as M3 type file), describing the real trajectory updated by radar tracks. 

The other source of information is given by the NEVAC files. NEVAC files \cite{NEVAC} contain the definition (borders, altitude, relationships, time of opening and closing) of airspace elements, namely airblocks, sectors, Flight Information Region (FIR), etc. The active elements at a given time constitute the configuration of the airspace at that time. Thus, they give the configuration of the airspaces for an entire AIRAC cycle. Here we only use the information on sectors, FIRs or ACCs and configurations to rebuild the European airspace. 

Our study investigates air traffic management on the en-route phase. After the filtering procedure 35704 flights are retained in the investigated AIRAC. In order to include the local constrains of the sector capacities, sectors are defined as described in reference \cite{Lillostrat}. These sectors are static bi-dimensional projection of the sectors higher than FL 350. 

%%%%%%%%%%%%%%%%%%%%%%%%%%%%%%%%%%%%%%%%%%%%%%%%%%%%%%%%%%%%%%%%%%%%%%%%%%%%%%%%%%%%%%%%%%%%%%%%%%%%
%%%%%%%%%%%%%%%%%%%%%%%%%%%%%%%%%%%%%%%%%%%%%%%%%%%%%%%%%%%%%%%%%%%%%%%%%%%%%%%%%%%%%%%%%%%%%%%%%%%%
%%%%%%%%%%%%%%%%%%%%%%%%%%%%%%%%%%%%%%%%%%%%%%%%%%%%%%%%%%%%%%%%%%%%%%%%%%%%%%%%%%%%%%%%%%%%%%%%%%%%
\section{The Model} \label{model}
%%%%%%%%%%%%%%%%%%%%%%%%%%%%%%%%%%%%%%%%%%%%%%%%%%%%%%%%%%%%%%%%%%%%%%%%%%%%%%%%%%%%%%%%%%%%%%%%%%%%
%%%%%%%%%%%%%%%%%%%%%%%%%%%%%%%%%%%%%%%%%%%%%%%%%%%%%%%%%%%%%%%%%%%%%%%%%%%%%%%%%%%%%%%%%%%%%%%%%%%%
%%%%%%%%%%%%%%%%%%%%%%%%%%%%%%%%%%%%%%%%%%%%%%%%%%%%%%%%%%%%%%%%%%%%%%%%%%%%%%%%%%%%%%%%%%%%%%%%%%%%

The agents of our agent-based model are aircraft/pilots and air traffic controllers who are active within an ACC in the European airspace. In the considered ACC all sectors are controlled from the same room (control center). In our ABM we simulate the realization of the en route trajectory starting from the planned trajectory.

The interaction between the aircraft/pilots and air traffic controllers is needed in order to manage the changes from the planned flight necessary due to unforeseen events, e.g. weather events, conflict resolutions, etc. Another important task involves the issuing of directs. A direct is a change of trajectory from the planned one such that the aircraft is allowed to follow a shorter flight trajectory in a region of the ACC by skipping one or more navigation points of the M1 flight-plan. In fact, whenever possible, the model allows directs that are given within an ACC in order to speed up the passage of the aircraft, provided that no additional conflict is created.

%%%%%%%%%%%%%%%%%%%%%%%%%%%%%%%%%%%%%%%%%%%%%%%%%%%%%%%%%%%
\subsection{Overview of the Model} \label{schemeA}
%%%%%%%%%%%%%%%%%%%%%%%%%%%%%%%%%%%%%%%%%%%%%%%%%%%%%%%%%%%

\begin{figure}  [H]
 \centering
                         \includegraphics[width=12cm]{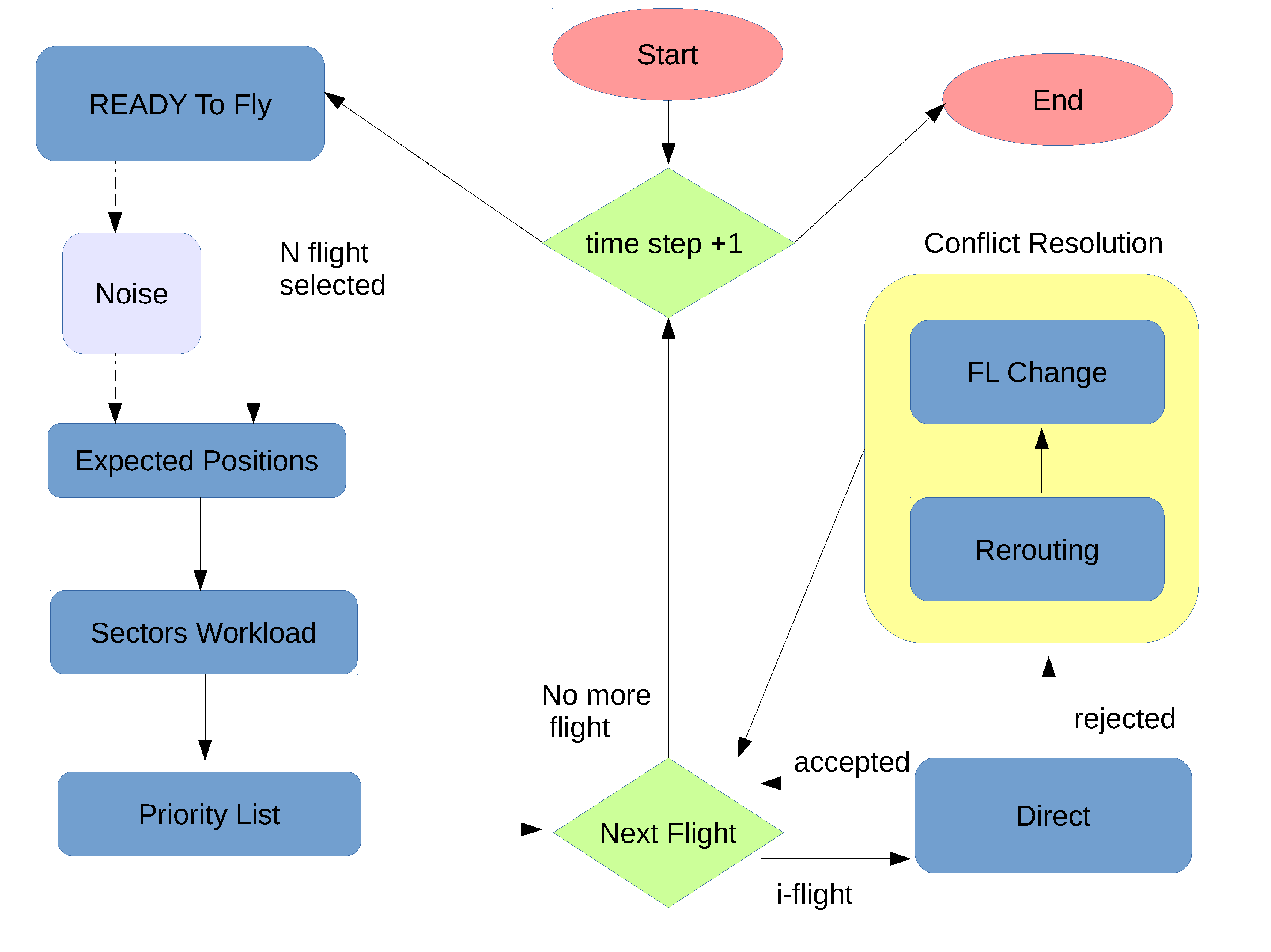} 
                         \caption{Schematic block diagram of the tactical ABM model.}\label{blockdiag}
\end{figure}

A schematic block diagram of the model is given in Fig. \ref{blockdiag}. Below we describe the different modules of our model.

We have designed the code in a modular way that allows to swap the priority of the strategies adopted by the controllers. In fact, as a default controllers first check for the possibility of doing re-routings and then change the flight altitude (FL change). Therefore, due to the modularity of the code the sequence among the different modules can be changed by the user of the code.

The code that implements the model presented here is written in C \cite{C} and it is available at the following URL: ${\tt{http://ocs.unipa.it/software.html:ELSA \, Tactical \, Layer}}$\footnote{A previous version of the code specifically dedicated for performing simulations in the SESAR scenario is available at the following URL: ${\tt{http://ocs.unipa.it/software.html:ELSA \, SESAR \, Simulator}}$}.

%%%%%%%%%% %%%%%%%%%%%%%%%%%%%%%%%%%%%%%%%%%%%%%%%%%%%%%%%%%
\subsection{Navigation Points} 

The planned flight trajectories are sequences of specific points of the airspace called navigation points to be crossed by the aircraft at specific times, flight levels, and within a specific sector. The velocity of each aircraft during the flight interval between two navigation points is assumed to be constant and its value is estimated from the schedule of the flight plan. In our simulations all navigation points present in the last-filed flight-plans are used. When trajectory changes are required by the controllers these changes involves temporary navigation points that are selected by the program or that can be externally provided. 
In the simulations we present in this paper the temporary navigation points are randomly uniformly distributed within the Italian ACC.

It should be noted that not all temporary navigation points will be used in the flight deviations. Only a set of them will be selected, as we will explain below. All the not used ones will be eliminated from the analysis after all the flights in the ACC will be managed. As we will explain in section \ref{rerout}, they are generated to allow the aircraft to deviate from the planned trajectories without necessarily passing over a predefined navigation point which might be too far.

%%%%%%%%%%%%%%%%%%%%%%%%%%%%%%%%%%%%%%%%%%%%%%%%%%%%%%%%%%%
\subsection{Timing and trajectory forecast of the model}  \label{timestepchoice}

The ABM is a discrete-time model. At time $t_0$ suppose that the position of all aircraft is known. The elementary time step of the model is $\delta t$. At time $t$ the time evolution of the system is computed with $\delta t$ time resolution until the time $t+\Delta t_l$ where $\Delta t_l$ is the look-ahead time of air traffic controllers. On the basis on the estimated time evolution air traffic controllers release their decisions to the aircrafts and a new iteration starts. To minimize the computational cost of the simulation the initial time of the next iteration of the model is performed at time $t+\Delta t_s$ where $\Delta t_s$ is a time interval ranging between $\delta t$ and $\Delta t_l$. The values used in our simulations after calibration were $\delta t=10$ seconds, $\Delta t_l=7.5$ minutes and $\Delta t_s=3$ minutes. 

In addition to the above time intervals the model uses a specific look-ahead when the emission of a direct is considered. We label this specific look-ahead interval as $\Delta t_d$. In our model $\Delta t_d$ and $\Delta t_l$ are integer multiple of $\delta t$.
\begin{figure}  [H]
\centering
                         \includegraphics[width=4cm]{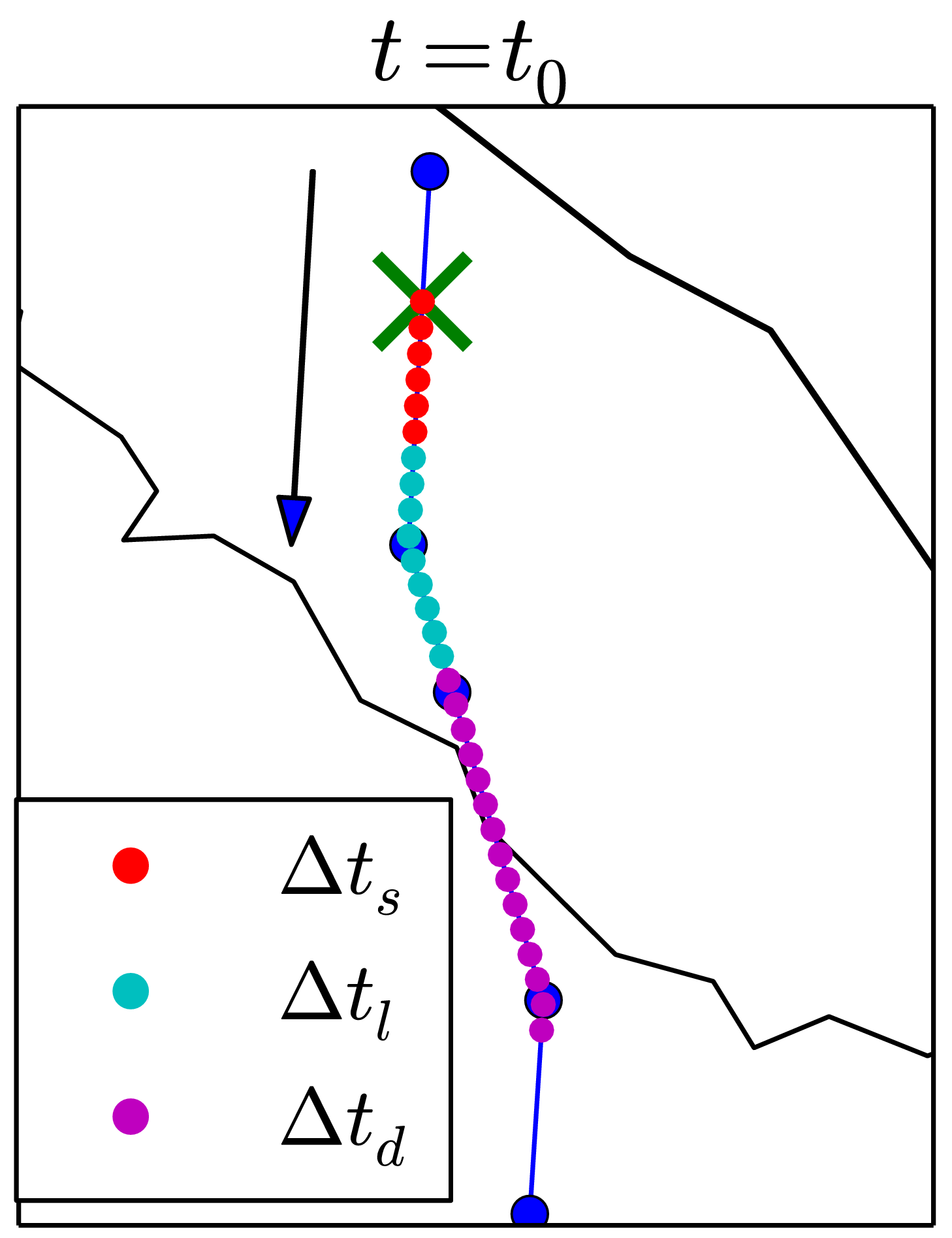} 
                         \includegraphics[width=4cm]{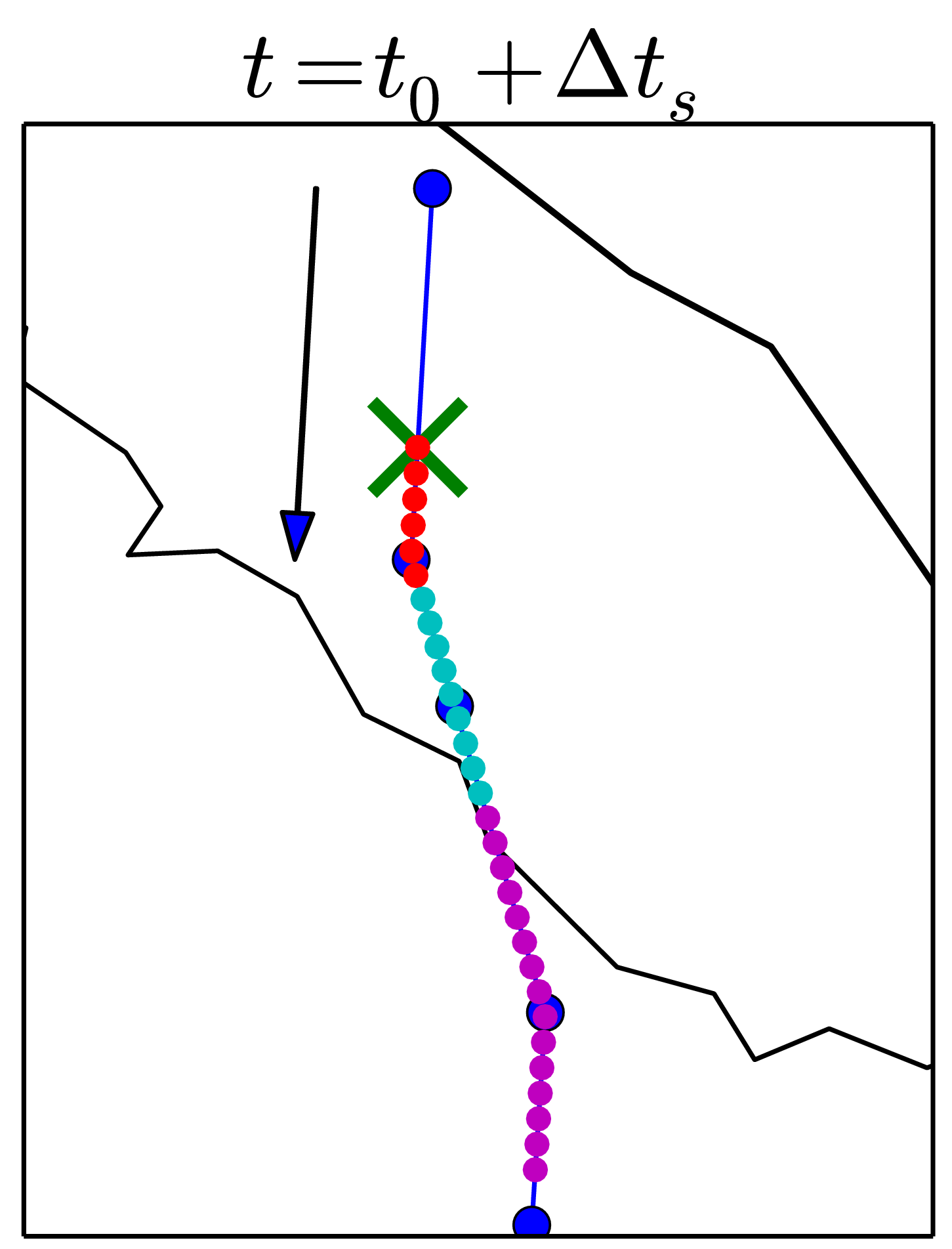}                          
                         \caption{Illustration of the time discretization used in the model. In the two panel we show a flight trajectory sampled at a discrete times with an elementary time interval of $\delta t=30$ sec. The dots indicate the aircraft positions sampled at each $\delta t$. Red dots are the aircraft positions evaluated within a time step $\Delta t_s=3$ min for the flight trajectory evaluated at $t=t_0$ (left panel) and at $t=t_0+\Delta t_s$ (right panel). Red and cyan dots are the aircraft positions evaluated within the time interval of the look-ahead $\Delta t_l=7.5$ min whereas red, cyan and magenta dots are the aircraft positions evaluated within the time interval of the look-ahead used to issue directs $\Delta t_d=15$ min. Blue circles are navigation points of the flight plan. The cross indicates the initial position of the aircraft at the initial time of the time step. The arrows indicate the directions travelled by the aircraft.}\label{fig:time_discr}
\end{figure}

%%%%%%%%%%%%%%%%%%%%%%%%%%%%%%%%%%%%%%%%%%%%%%%%%%%%%%%%%%%
%\subsubsection{Velocity Noise Module}  \label{velnoisemod}

Indeed, in the basic setup, the controller forecast of the aircraft position is exact within its time look-ahead. Our ABM allows to introduce some errors in the forecast of the controller. This is done by setting a parameter $l_\epsilon \ne 0$ which is controlling the uncertainty in the estimation of the velocity of the aircraft. Specifically, the uncertainty in the controller's forecast is introduced by the following procedure:
 (i) between time $t$ (current time) and $t+\Delta t_s$, the model that the aircraft maintains the planned velocity, (ii) between $t+\Delta t_s$ and $t+\Delta t_l$, the model introduces an uncertainty in the aircraft velocity the velocity. The velocity used by the controller is  $v (1+\epsilon_v)$, where $\epsilon_v$ is drawn at random from a uniform distribution in the range $-l_\epsilon$ and $l_\epsilon$. 
With this choice the controller makes bigger errors on positions on longer times. 

It should be noted  that the actual velocity of aircrafts do not change in our model, and hence they are always those of the planned trajectories (except in case of re-routing where the velocity is extrapolated on the new segment). In practice and in the absence of learning processes, incorrect forecasts or stochastic changes of the trajectories are indistinguishable.

%%%%%%%%%%%%%%%%%%%%%%%%%%%%%%%%%%%%%%%%%%%%%%%%%%%%%%%%%%%
\subsection{Priority list of controllers' actions} 

At each time step we create a list $FL$ of flights active in the considered time-step. The list is randomly ordered. The order of the list is followed by the controller in her attempt to solve potential conflicts and to issue directs. Specifically, the $i$-th aircraft trajectory in the list is checked against the trajectories of previous listed $i-1$ flights. For example, the first aircraft in the list will perform its planned trajectory whereas the trajectory of the second one will be checked with respect to the trajectory of the first one. The trajectory of the third aircraft will be checked with respect to the trajectories of the second and the first ones and so on. Indeed, to speed computation, the trajectory check between two aircrafts is not performed when the two trajectories are too far to interact within the look-ahead time interval. 

The random reordering of the flight priority list is done in order to be sure that the trajectories to be deviated are not always the same ones. If a conflict involving the $i-th$ aircraft is not solved by one of the procedures followed by the controller, then the list is modified by putting the $i-th$ aircraft in the first position of the list and the trajectory analysis of the time step is repeated from the beginning. When this redefinition of the priority list is repeated more than 50 times for a time step the simulation is aborted.

%%%%%%%%%%%%%%%%%%%%%%%%%%%%%%%%%%%%%%%%%%%%%%%%%%%%%%%%%%%
\subsection{Conflict Detection module} \label{deconfmodule}

The collision detection module calculates the minimum distance for each pair of aircrafts positions between the flight $i$ and the flights labeled as $i-1$ in the priority list. This operation is repeated for all the times $t+k \delta t$ with k ranging from 1 to N such that $t+N \delta t \, \equiv \, t +\Delta t_m$ where $\Delta t_m$ is equal to $\Delta t_l$ or $\Delta t_d$ depending whether the conflict detection module has been activated by a re-routing procedure or by a direct procedure.   
For each flight $i$ the algorithm computes an array of flight positions ${p}_{i,k}$, $k=1, \cdots N$ given the flight positions at different time $t+k \delta t$.

Suppose we are now checking if the $i$-th flight trajectory is conflicting with all other $f_j$ trajectories, with $j < i$. For each of the $N$ elementary time-increments, we compute a matrix of distances $d^i_{j,k}$ with j rows and N columns. 
For all aircrafts flying at the same flight level all distances are computed by using the Haversine distance \cite{haversine} between each pair of flight positions\footnote{The computation of the Haversine distance is particularly time-consuming. Therefore we have also implemented in the code the possibility that in some specific cases the Euclidean distance is used instead of the Haversine one. This is for example advised when it is necessary to perform a very large number of simulations in a limited portion of the airspace.}. For pairs of aircrafts flying at different flight levels at time $t+k \delta t$ the distance is set to infinity because aircrafts flying at different flight level are not raising minimum separation issues. For each column we select the minimum value and obtain a vector $d^i_{min}(k)$ of length $N$. 
A possible conflict between two aircrafts flying at the same flight level is detected at time $t+k \delta t$ whenever the elements of $d^i_{min}(k)$ are smaller than the safety distance threshold $d_{thr}$ that is usually set to 5 NM. This reference value is the standard value used in ATM for conflict detection.

In order to mimic some heuristics typically used by air traffic controllers in detecting conflict  we introduce in the ABM a linear growth of the safety threshold $d_{thr}$ as a function of the time interval from the present time.  In fact when an air traffic controller forecasts the position of an aircraft at a far future time  he uses an additional space of separation between the aircraft to be safe in the forecast. Our model therefore uses a safety distance threshold defined as:
\begin{eqnarray}
                           d_{thr} (k) =  d_{thr} + \Delta d_{thr} k
\end{eqnarray}
where $\Delta d_{thr}$ is one of the model parameters.

 When a conflict is detected the algorithm proceeds to the next module that performs the de-conflicting of flight trajectories.

%%%%%%%%%%%%%%%%%%%%%%%%%%%%%%%%%%%%%%%%%%%%%%%%%%%%%%%%%%%
\subsection{Conflict Resolution module} \label{TACTmod}

After the conflict detection module has detected a conflict, this module searches for a new conflict free trajectory. It is conceived as a two-step algorithm that acts on the search of a new trajectory. The first step attempts to perform a re-routing of the flight trajectory. When the re-routing is successful the new trajectory is accepted. If the re-routing module fails to find an appropriate new trajectory the algorithm move to the second step that require a change of flight level for the aircraft.

%=======================================
\subsubsection{Re-reouting submodule} \label{rerout}

The procedure of the re-routing attempt is illustrated in Fig. \ref{fig:re-routing}. We first identify the position $B$ (not necessarily a navigation point) defined by $k=0$ at the considered time step.  We then identify the navigation point $A$ which is the first navigation point after the area of the potential collision (filled circle in the figure). The procedure is to attempt to re-route the trajectory such that all navigation points that are in the conflict area plus the $A$ navigation point are avoided. These navigation points are replaced by a temporary navigation point (see T point in Fig. \ref{fig:re-routing}). The temporary navigation point is selected from several possibilities (see grey points  in Fig. \ref{fig:re-routing}) by choosing the navigation point solving the conflict that presents the shortest path  between position B and navigation point E, i.e. the navigation point where the flight trajectory is re-routed. Another constraint about the re-routing trajectory is the request that the deviated trajectory from the planned one cannot exceed an angle $\alpha_M$ both for the $\alpha_{in}$ and $\alpha{out}$ angles observed between the planned and the re-routed trajectories (see Fig. \ref{fig:re-routing}). 
If the re-routing trajectory is not able to find a solution the re-routing submodule attempts to re-route the flight trajectory by moving forward the navigation point E and by looking again for a re-routing trajectory. When a possible solution is found, the result of the search is accepted if the re-routing trajectory deviates from the planned trajectory for less than a maximal time $T_{max}$. If the solution found has a deviation time longer than $T_{max}$  the re-routing submodule is not selecting any new trajectory and the resolution of the conflict is passed to  the flight level module. {In the right panel we show the distance between the two aircraft for the planned trajectory (blue dots) and the ones considered by the ABM module (gray lines). Amongst those, the trajectory that satisfies the requirement of minimum separation distance is highlighted in red, as in the left panel.
\begin{figure}
 \centering
                         \includegraphics[width=8cm]{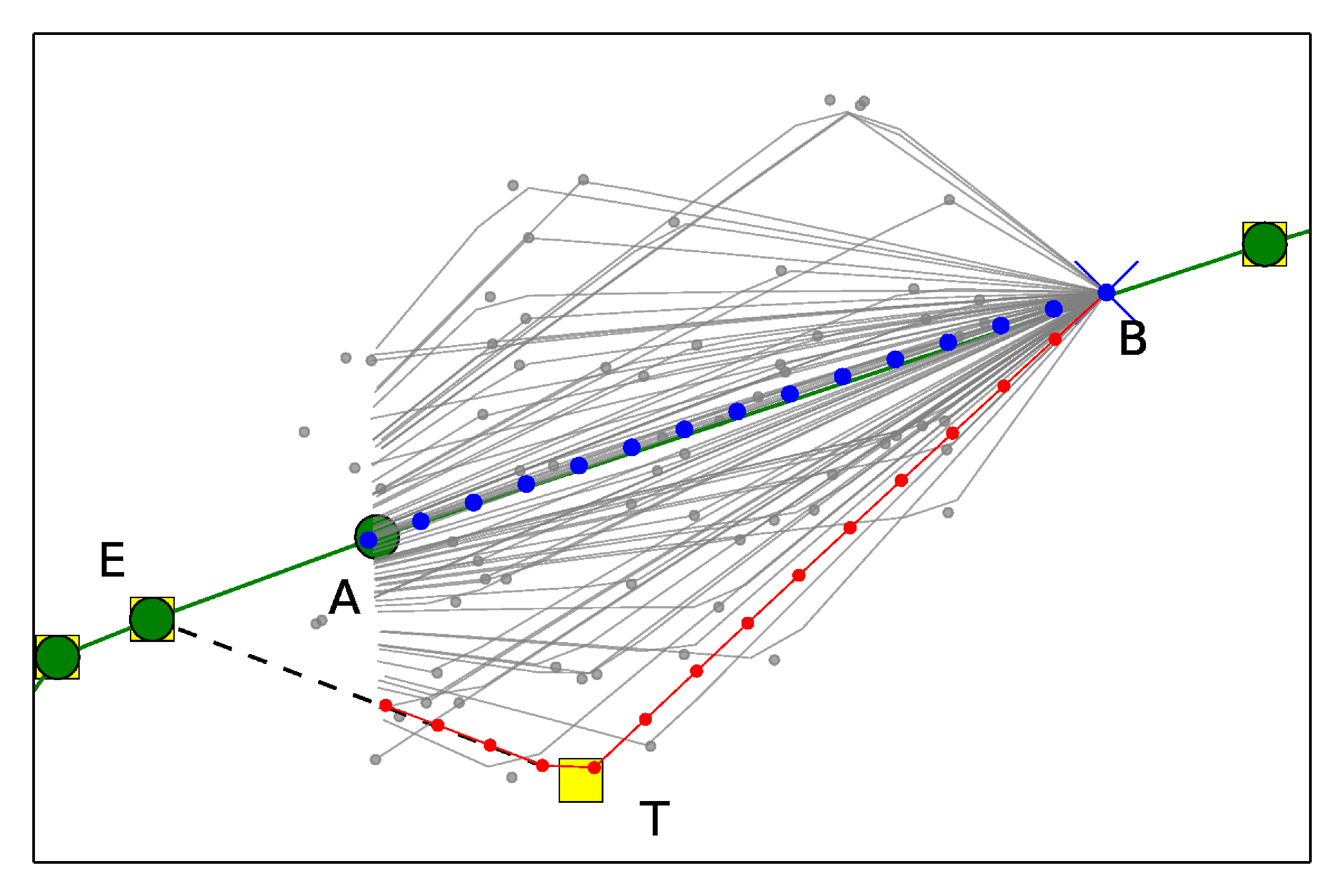} 
                          \includegraphics[width=8cm]{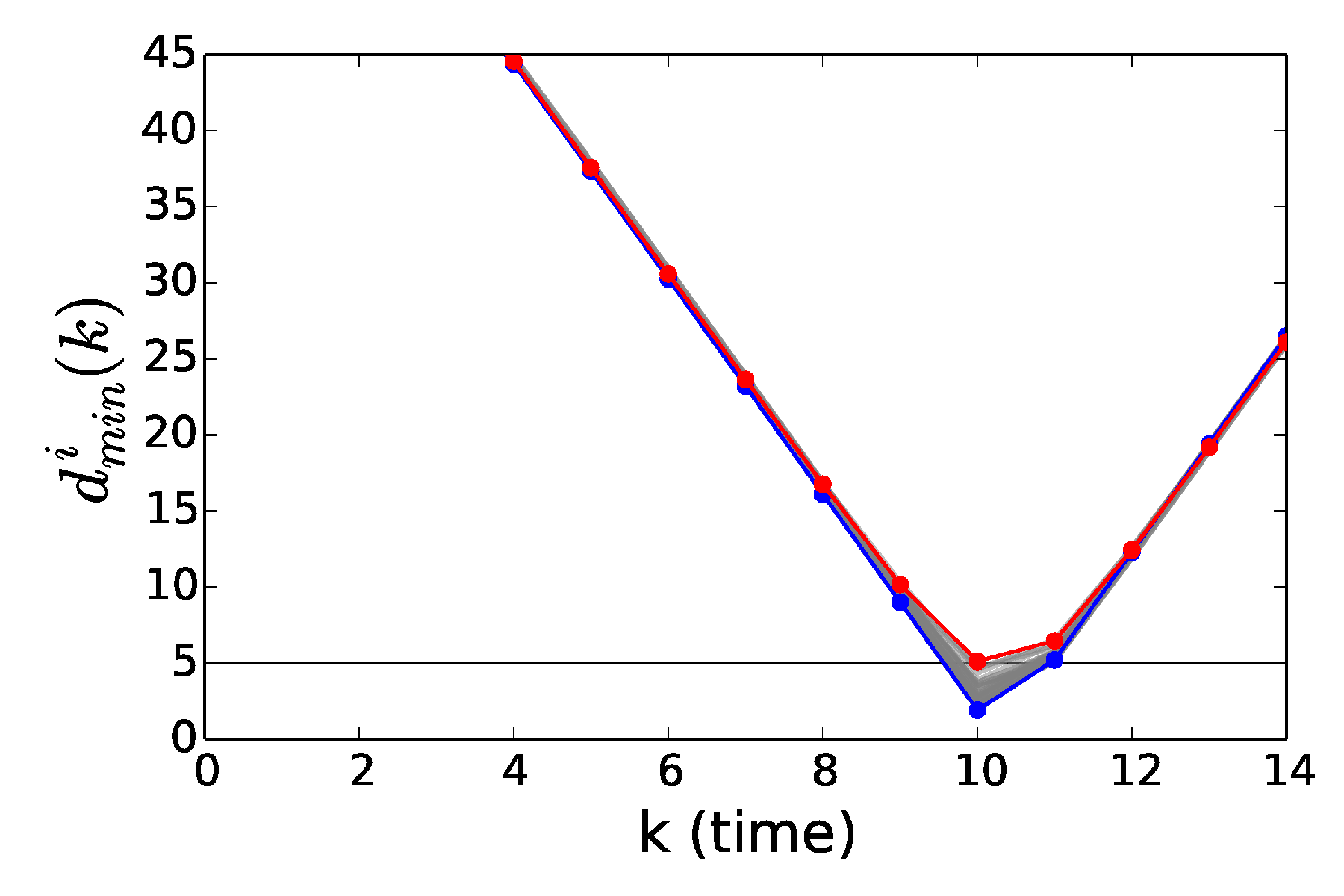}                   
                         \caption{The figure illustrates the procedure of re-routing, see text for more details. The gray trajectories, although possible, are not selected because they do not guarantee the minimum separation of 5 NM required between two aircrafts. The re-routing occurs between point B (blue cross of the left panel) and E (green circle of the left panel).  The re-routing is performed by deviating the flight trajectory to the temporary navigation point $T$ (yellow square of the left panel) and then re-route back the trajectory to navigation point E. To find the best re-routing flight trajectory the ABM module explores trajectories passing through several different temporary navigation points (gray spots of the left panel). The distance between the two aircrafts is shown in the right panel for the planned trajectory (blue dots), the ones considered by the ABM module (gray lines) and the selected one satisfying the requirement of minimum separation distance (red dots). In the left panel all trajectories are considered within the time of the look-ahead $\Delta t_l$.}\label{fig:re-routing}
\end{figure}

%=======================================
\subsubsection{Flight level change submodule} 

The second step of the conflict resolution module involves changes of flight level. A flight level (FL) is a unit measure defined as altitude above sea-level in 100 feet units measured according to a standard atmosphere. Allowed flight levels are separated by 1000 feet, i.e. 10 flight levels (separation levels). This is the standard separation vertical distance between any pair of aircraft. Moreover, in our model the semicircular rule has been considered, meaning that aircraft flying in opposite directions are allowed to fly only along odd or even levels respectively. Therefore when an aircraft needs to be moved to another separation level, it will not be moved to the next first one but to the second one in order to respect the semicircular rule, thus performing a jump of 2000 feet or 20 FLs.

All flights are considered to be available in the planned trajectories. In our agent based model aircrafts can move two Flight Levels (FL) first upwards and, if the conflict cannot be solved by a move upwards,  downwards.  The model assumes that the flight level change is abrupt occurring when the conflict resolution is settled. If no flight level is available to solve the conflict then the list is reshuffled by moving the considered flight in the first position of the priority list. 

When a flight level change is executed the flight remains in the new flight level for a time equals to $T_{max}$. After $T_{max}$ the aircraft goes back to the flight level of the planned flight.

%%%%%%%%%%%%%%%%%%%%%%%%%%%%%%%%%%%%%%%%%%%%%%%%%%%%%%%%%%%
\subsection{Direct module} \label{directsD}

A direct, i.e. a change of the planned trajectory significantly shortening the flight path, is made by skipping one or more navigation points of the flight plan and flying straightly from the current navigation point to a distant navigation point of the flight plan. In our algorithm this module is executed with a probability that depends on the workload of sectors of initial and ending navigation points of the direct. 

%=======================================
\subsubsection{Sector workload} \label{multisectormodule}

The ACC we are considering is divided into a number of sectors. Each sector is characterized by its geographical location and by a proxy for its capacity, defined by us as inferred capacity and estimated  as the maximum number of aircrafts that are simultaneously present in a sector within a time window of one hour \cite{Lillostrat}. This information is obtained from the flight plans of the AIRAC used to start simulations.

In addition to the inferred capacity of a sector we dynamically estimate its workload. Specifically, we estimate sector workload by assigning a numerical flag to each navigation point of the planned flight trajectories for each sector.  We define workload of a sector the number of flights planned to cross it during the time window of an hour.  At each time-step the ABM evaluates the workload of each sector of the ACC. 

When the workload of sector exceeds its inferred capacity all directs that come from other sectors are not allowed, while re-routing due to safety issues are still allowed. Operatively this means that in a condition when the workload equal or exceeds inferred capacity any other incoming flight has to enter the sector from the navigation point present in the flight plan.

%=======================================
\subsubsection{Direct execution} \label{dexecmodule}

Specifically, let $n_i$ be the first navigation point to be crossed of the current time step, and $n_m$ the navigation point where the flight will return on its original flight plan. By issuing a direct trajectory from $n_i$ to $n_m$  therefore $m-i-1$ navigation points will be absent in the new trajectory, as illustrated in the left panel of Fig. \ref{fig:Art_Direct}.

The direct module first evaluates how many navigation points can be skipped with the constraint that the flight has to come back to the planned trajectory within a time interval equal to $T_{max}=20$ min \footnote{The choice of $T_{max}$ has been done in agreement with the indications of the air traffic controllers consulted within the ELSA project of SESAR.}, and the direct is conditioned to the inferred sectors' capacity of the adjacent sectors. 

After that the model evaluates if the new trajectory will be involved in conflicts. In order to do this check we use the Conflict Detection module of section \ref{deconfmodule} with a different time-interval $\Delta t_d$.  If the direct is safe and the angle between the new and original trajectory is larger then a sensitivity threshold value $\alpha_s=1^\circ$ then the new trajectory is accepted, otherwise the algorithm tries a suboptimal solution, see the left panel of Fig. \ref{fig:Art_Direct}.
\begin{figure}[H]
\centering
                 \includegraphics[width=8cm]{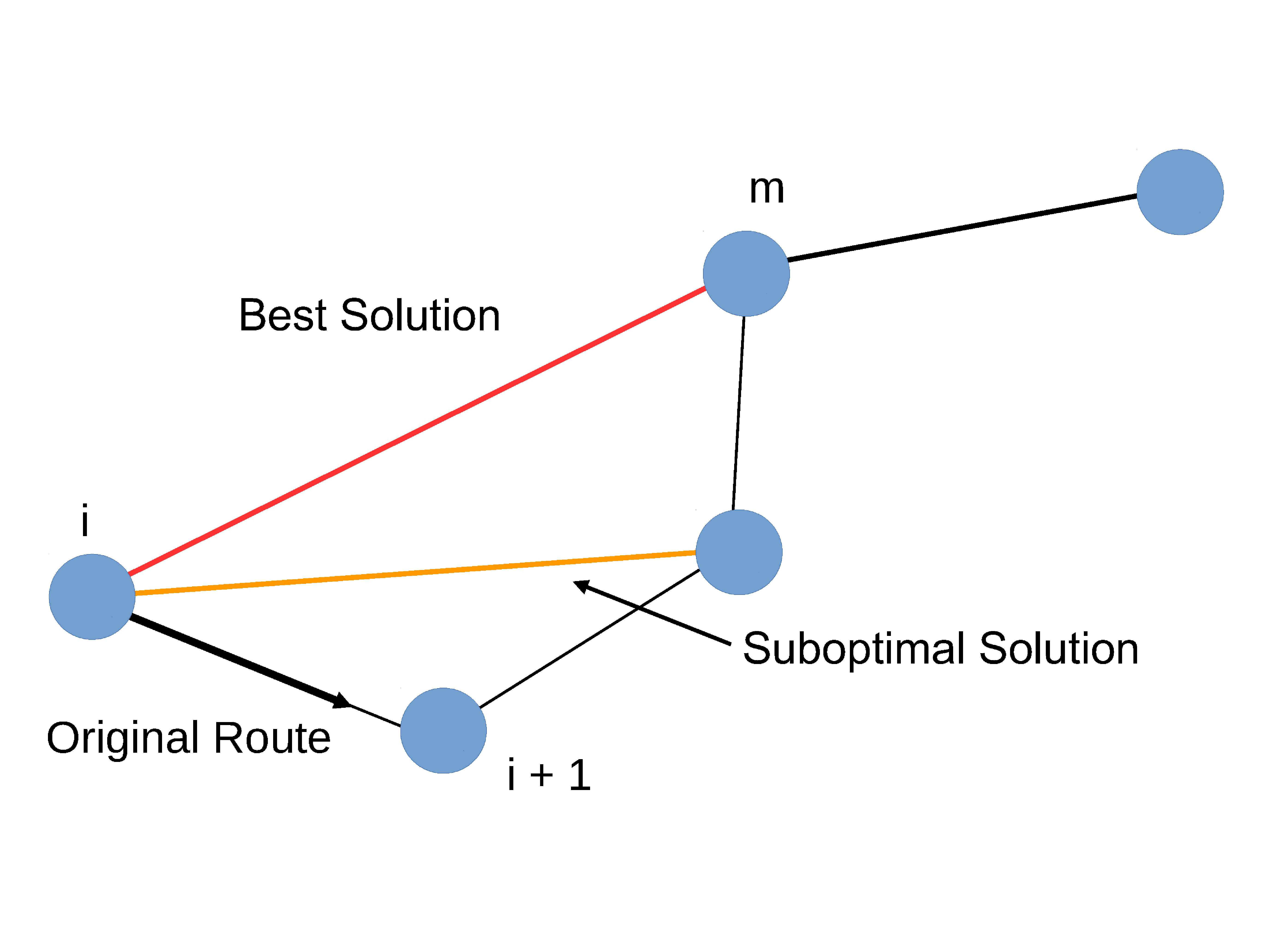} 
                 \includegraphics[width=8cm]{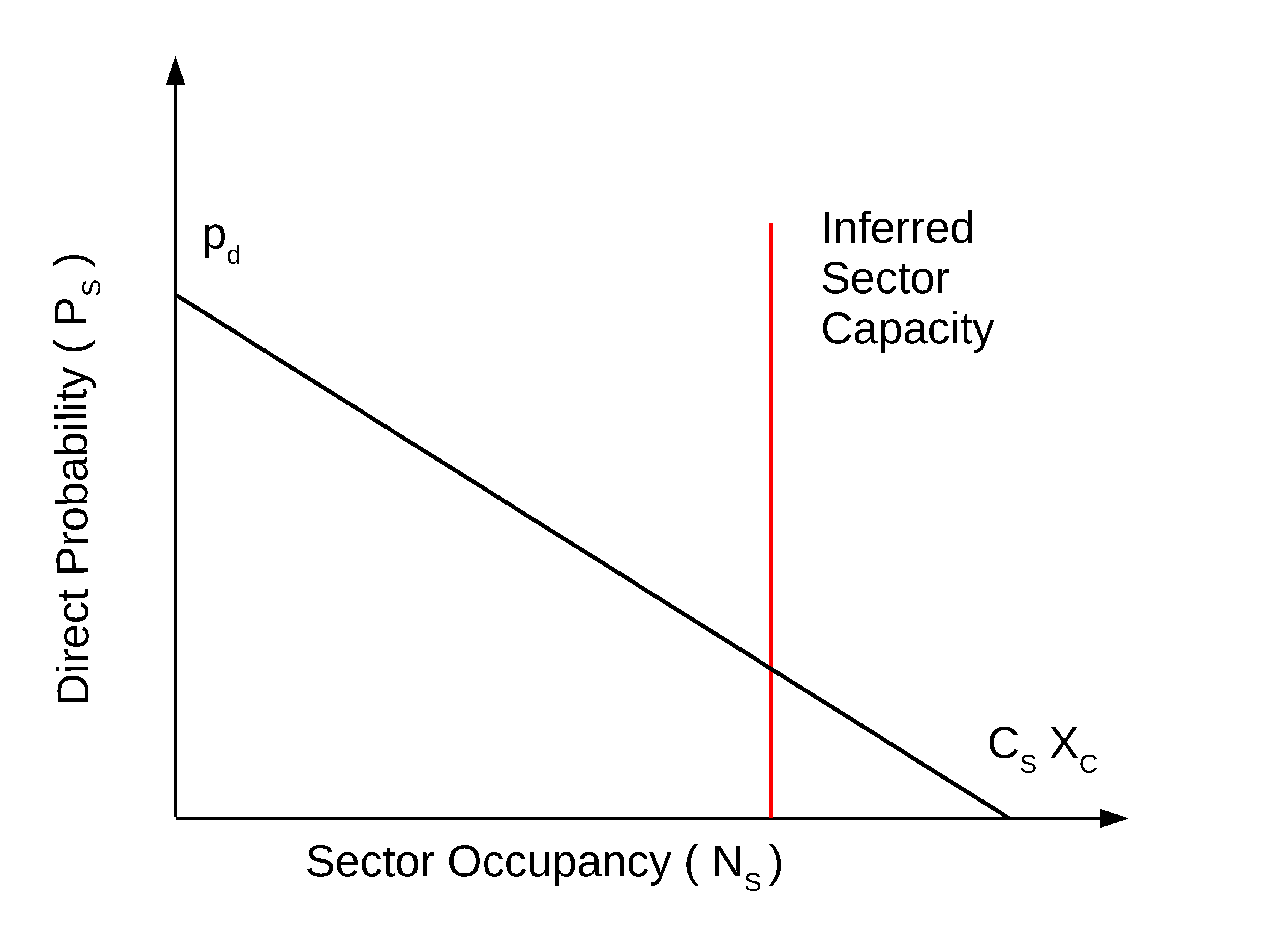} 
                 \caption{Left panel: illustration of the procedure implemented to issuing directs. Right panel: Probability function used in the procedure implemented to issuing directs.}\label{fig:Art_Direct}
\end{figure}

The probability to issue a direct for an air controller operation on a sector $s$ is dependent by the workload and by the inferred capacity of the sector involved. Let $C_s$ be  a constant of the $s$-sector that in our calibration procedure we fixed to be the inferred sector capacity obtained from real data \cite{Lillostrat}. Let $P_s(N_s)$ be the probability to issue a direct in the $s$-sector when the workload of sector $s$ is $N_s$. For the sake of simplicity we model $P_s(N_s)$ as a linear decreasing function of $N_s$, see the right panel of Fig. \ref{fig:Art_Direct}
\begin{eqnarray}
                          P_s(N_s)=p_d \, \Bigl(
                                                              1 - {N_s \over x_c \, C_s} \label{dirprob}
                                                     \Bigl)
\end{eqnarray}
The probability to attempt a direct is function of two parameters $p_d$ and $x_c$. The first $P_s(N_s=1) = p_d$ is the probability to attempt a direct if just one flight is in the sector. The second parameter $x_c$ is used to control the slope of probability as a function of the workload, as illustrated in the right panel of Fig. \ref{fig:Art_Direct}. The $p_d$ parameter plays the role of a scale factor for the overall probability. The $x_c$ parameter measures the controllers confidence in approaching the maximum sector's inferred capacity. While $N_s$ and $C_s$ are parameters depending on each specific sector, $p_d$ and $x_c$ are global parameters that are set across the whole considered ACC. 

In the present version of the model, air traffic controllers behave in the same way in the different sectors. However, by introducing a direct probability $P_s$ that depends on the actual inferred capacity of each sector, see Eq. \ref{dirprob}, we have realized a genuinely multi-sector ABM where directs are issued differently across the ACC and across the day. The choice of the use of the same parameters for different controllers and sectors (except inferred capacity) is done in order to make the ABM model as parsimonious as possible.

%%%%%%%%%%%%%%%%%%%%%%%%%%%%%%%%%%%%%%%%%%%%%%%%%%%%%%%%%%%
\subsection{Model's parameters} \label{summodpar}

In Table \ref{modpar} we summarize the model's parameters used in the different modules described above. In the third column of the Table we give a short description of the parameter and in the fourth column we give a categorization of the parameters describing whether the parameter is calibrated from data (CD) or it is set according to information obtained by interviewing ATM experts and ATCOs (CV). 

The parameters that need to be calibrated from data are a few. There are also several parameters (CV category) that can be inferred form the typical behavior of controllers.  These are parameters that should be selected by consulting ATM experts and ATCOs. It is worth noting that by considering these variables as parameters our model allows to perform scenario simulations to test how changing a certain feature of air traffic controllers might affect ATM performances.

\begin{table}[H]
\centering
\begin{tabular}{|c|c|l|l|c|}
\hline
{\bf{ID}} & {\bf{Parameter}}  & {\bf{Description}}  & {\bf{Type}} & {\bf{Value}}\\
\hline
1 & $\delta t$ & Length of the elementary time-interval.  & ~ &  10 s \\
\hline
2 & $\Delta t_l$ &  Length of the forecast for collision, controller's look-ahead.  & CD & 7.5 m  \\
\hline
3 & $\Delta t_d$ &  Length of the forecast for directs.  &  CV & 15 m  \\
\hline
4 & $\Delta t_s$ & Basic time step.  & ~ & 3 min. \\
\hline
5 & $l_\epsilon$ & Aircraft velocity noise range & ~  & 0 ; 0.1\\
\hline
6 & $d_{thr}$  & Safety Distance threshold. & CV & 5NM \\
\hline
7 & $\Delta d_{thr}$ & Increment of the safety distance threshold in the forecast. & CV & 0; $0.33$ in section \ref{DST}\\
\hline
8 & $\alpha_M$ &  Maximum angle of deviation between planned & & \\
   &                     &   and re-routed trajectory.  & CV & 27 deg. \\
\hline
9 & $T_{max}$ &  Maximum time spent outside  & &\\
   &                    &  the planned flight trajectory.  & CV & 20 min.\\
\hline
10 & $p_d$ & Unconditional probability to try to issue a direct. & CD & 0.24 \\
\hline
11 & $x_c$ & Tolerance to the sector congestion. & CD & 0.63 \\
\hline
12 & $\alpha_s$ & Minimal angle to issue a direct.  & ~ & 1$^\circ$ \\
\hline
13 & $C_s$ & Sector inferred capacity.   & CD &  Sector's specific \\
\hline
\end{tabular}
\caption{Model parameters.}  \label{modpar}
\end{table}

%%%%%%%%%%%%%%%%%%%%%%%%%%%%%%%%%%%%%%%%%%%%%%%%%%%%%%%%%%%%%%%%%%%%%%%%%%%%%%%%%%%%%%%%%%%%%%%%%%%%
%%%%%%%%%%%%%%%%%%%%%%%%%%%%%%%%%%%%%%%%%%%%%%%%%%%%%%%%%%%%%%%%%%%%%%%%%%%%%%%%%%%%%%%%%%%%%%%%%%%%
%%%%%%%%%%%%%%%%%%%%%%%%%%%%%%%%%%%%%%%%%%%%%%%%%%%%%%%%%%%%%%%%%%%%%%%%%%%%%%%%%%%%%%%%%%%%%%%%%%%%
\section{Calibration of the model} \label{calibration}
%%%%%%%%%%%%%%%%%%%%%%%%%%%%%%%%%%%%%%%%%%%%%%%%%%%%%%%%%%%%%%%%%%%%%%%%%%%%%%%%%%%%%%%%%%%%%%%%%%%%
%%%%%%%%%%%%%%%%%%%%%%%%%%%%%%%%%%%%%%%%%%%%%%%%%%%%%%%%%%%%%%%%%%%%%%%%%%%%%%%%%%%%%%%%%%%%%%%%%%%%
%%%%%%%%%%%%%%%%%%%%%%%%%%%%%%%%%%%%%%%%%%%%%%%%%%%%%%%%%%%%%%%%%%%%%%%%%%%%%%%%%%%%%%%%%%%%%%%%%%%%

In this section we want to discuss the calibration activities that have to be performed in order to use our model. 
%We can distinguish two classes of parameters. Those that are relevant for the setting of the traffic conditions and those related to the different strategies implemented by the controllers in order to manage the aircraft trajectories.

We will here refer to the air-traffic of LIRR ACC (Rome, Italy) between 2010-05-06 and 2010-06-03, i.e. the 334 AIRAC. The input data of the model are taken from the database of DDR and NEVAC files developed within the ELSA \cite{D1p3} project and described in section \ref{datasec}. We consider as an input to the model the M1 flight plans with the constraints indicated in section \ref{datasec}.
%: 1) flights are performed with Landplanes (i.e. no helicopter, gyrocopter, only aircraft which can only operate from or alight on land), 2) flights are scheduled, 3) flights have a IATA code 4) flights have a duration longer than 10 minutes. 
To focus our attention on the en-route phase we filtered out from the flight plans all navigation points crossed at an altitude lower then 240FL. After the filtering procedure 35704 flights were retained in the entire AIRAC. In order to include the local constrains of the sector capacities, it is important to remember that the sectors are not static geometric regions but they are merged together and split dynamically  to fulfill the occupancy requirement. For the sake of simplicity we will refer to the collapsed sector defined in the reference \cite{Lillostrat}. These are a static bi-dimensional projection of the sectors higher than FL 350. The sectors capacity inferred from data is defined as the maximum number of flight expected within a time-window of a hour inside the collapsed sector.

In this section we describe our calibration procedure. In our simulations we consider the scheduled flights  of the LIRR ACC (Rome, Italy) of the AIRAC 334 described in section \ref{datasec}. The calibration procedure is performed by choosing a specific stylized fact observed in real data and requesting that model simulations are able to reproduce them.

Indeed, there is some degree of arbitrariness in selecting the specific stylized fact. Different ones can be chosen depending on the specific aspects of the ATM researchers want to investigate. In the present work, in order to calibrate the models parameters related to controllers' behavior we choose as stylized fact a statistical regularity concerning the intraday pattern of directs issued by ATCOs. Specifically we calibrate our model to reproduce the intraday evolution of the deviation rate metric that has been recently introduced in Ref. \cite{pap47}. 

The deviation rate introduced in  Ref. \cite{pap47} quantifies the deviations observed from the planned flight trajectories. We call {\em{deviation}} the event such that an aircraft passing over a scheduled navigation point does not go to the next planned one. The deviation rate is defined as the ratio between the observed number of deviations and the number of possible deviations in the airspace estimated in a one hour time window. The number of possible deviations is defined as the number of planned navigation points that are actually crossed by the aircraft. This metric is computed for each hour of the day by using the information about all planned and realized flight trajectories.

This metric describes an unknown mixture of ATCO operations, i.e. re-routing and direct. In \cite{pap47} it is shown that, in relative terms, directs are mainly issued during night-time i.e. in low traffic conditions while they are relatively less issued during day-time. Our choice is to reproduce this intraday statistical regularity. In the right panel of Fig. \ref{Fig.Calibration} we show (blue circles) the empirical behavior of the deviation rate estimated over the entire 334 AIRAC cycle as a function of the time of the day.  The deviation rate presents a U-shape having higher values during night hours and lower values during day hours.  The error bars are computed as the 95 \% Wilson score interval \cite{Wilson} used to associate a confidence interval to a proportion in a statistical population. 

Hereafter  we detail the procedure we have used to calibrate $p_d$, $x_c$  and $\Delta t_l$ parameters. 
In our calibration procedure we considered $p_d \in [0.03,0.5]$ with steps of 0.01567, $x_c \in [0.34,1.5]$ with step of 0.03867 and $\Delta t_l \in [5, 15]$ minutes with steps of 2.5 minutes and for each triplet of parameters we performed one single simulation for each considered day in the AIRAC, totaling 20 days of simulations -- with Saturday and Sundays excluded. From the output of the ABM we estimated the deviation rate with a time window of one hour. By using the results of simulations, we minimized the chi-squared $\chi^2$ computed starting from the deviation rates obtained with the ABM and the values estimated from real data. The $\chi^2$ is therefore computed over 24 points. In the left panel of Fig \ref{Fig.Calibration} we are showing the average values of the $\chi^2$, as a function of $p_d$ and $x_c$ when $\Delta t_l=7.5$ minutes. The lowest value of $\chi^2$ is associated to $p_d=0.2465$ and $x_c=0.6310$ and $\Delta t=7.5$ min. This set of parameters corresponds to $\chi^2=0.01294$. However, it is worth noticing that a larger region of parameters (see the magenta region) could still provide acceptable set of parameters. The solid green line in the right panel of Fig. \ref{Fig.Calibration} shows the deviation rate metric obtained by performing the simulation of the model with the selected parameters $p_d=0.2465$ and $x_c=0.6310$ and $\Delta t_l=7.5$ min. 
\begin{figure}[H]
\centering
	\includegraphics[width=8.5cm]{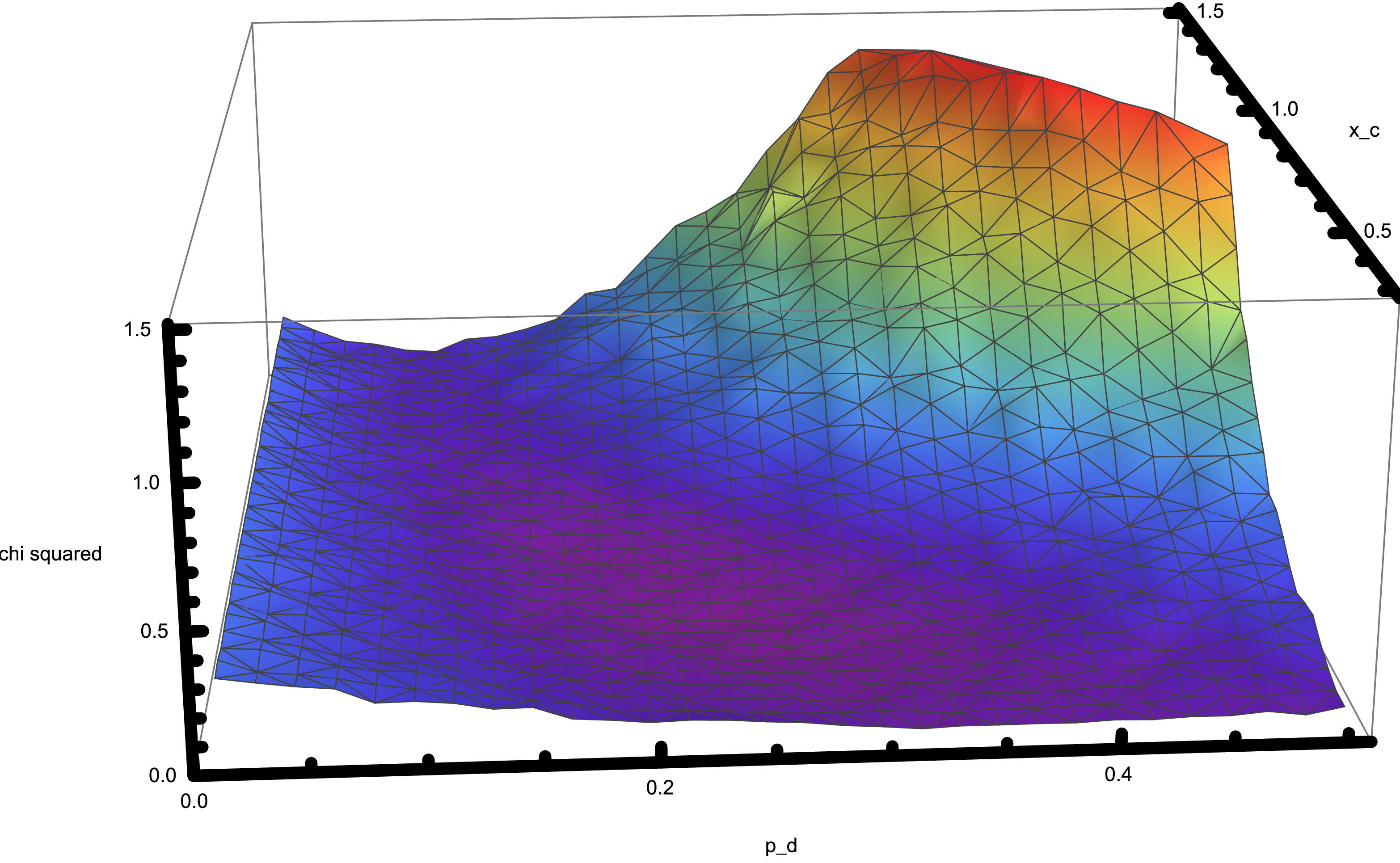} 
	\includegraphics[width=8.5cm]{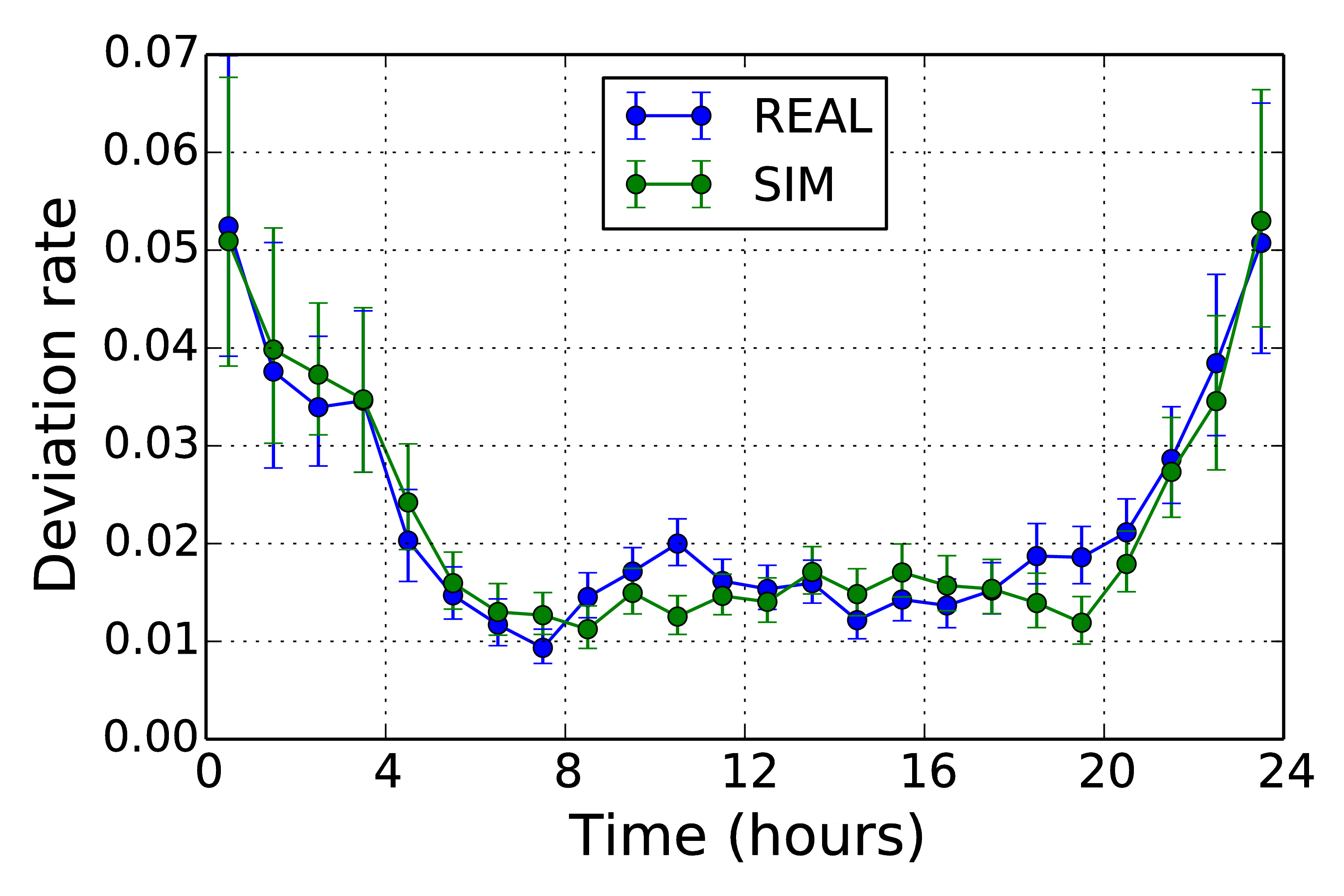} 
\caption{Illustration of the calibration procedure. Left Panel: we are showing the values of the $\chi^2$, as a function of $p_d$ and $x_c$ when $\Delta t=7.5$ minutes. Right Panel: we show the empirical (blue circles) behavior of the deviation rate metric averaged over the entire 334 AIRAC cycle. The solid green line shows the deviation rate metric obtained by performing a simulation of the model with the selected parameters $p_d=0.2465$ and $x_c=0.6310$ and $\Delta t=7.5$ min.}\label{Fig.Calibration}
\end{figure}

Here we want to assess the importance of the calibration procedure. In fact, in Fig.  \ref{fig:mappenocal} we show results that can be obtained by our model by choosing sets of parameters different from the calibrated ones. The first example sets that no direct is issued  (left panel of Fig.  \ref{fig:mappenocal} ). The ``No Directs'' case is obtained by setting $p_d=0$ and $\Delta t_l=7.5$ min.
The second example sets that the probability to issue a direct is independent from the sector workload  (right panel of Fig.  \ref{fig:mappenocal}).  This second example is obtained by setting to the case when $p_d=0.24$, $\Delta t_l=7.5$, as in the calibrated case and $x_c=1000$. Such a value of $x_c$ ensures that the sector workload plays no role when directs are issued. With the chosen parameters we have that the deviations rate simulated during night-time corresponds to the empirical case. 

\begin{figure}[H]
\centering
                 \includegraphics[width=7.5cm]{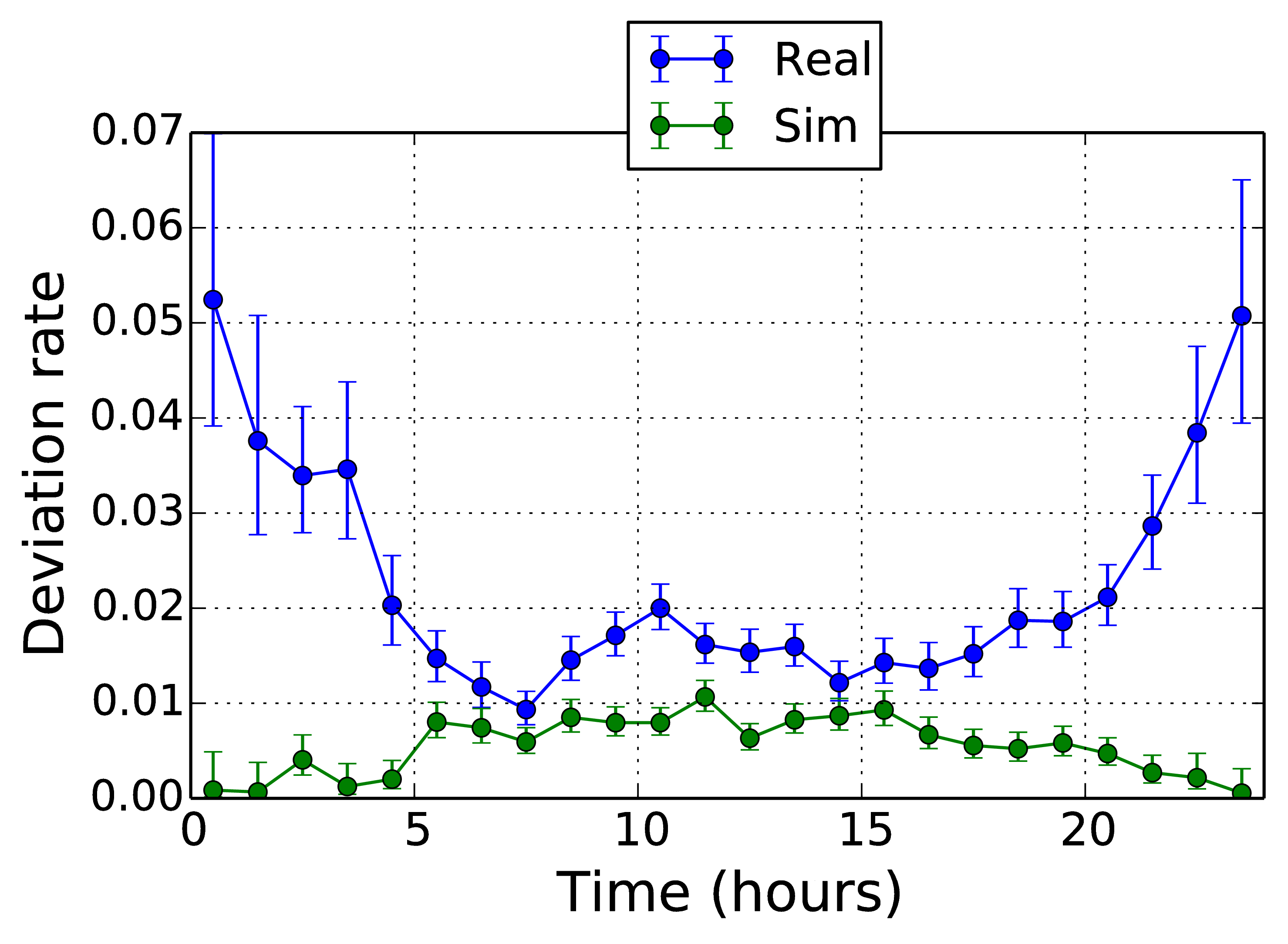} 
                 \includegraphics[width=7.5cm]{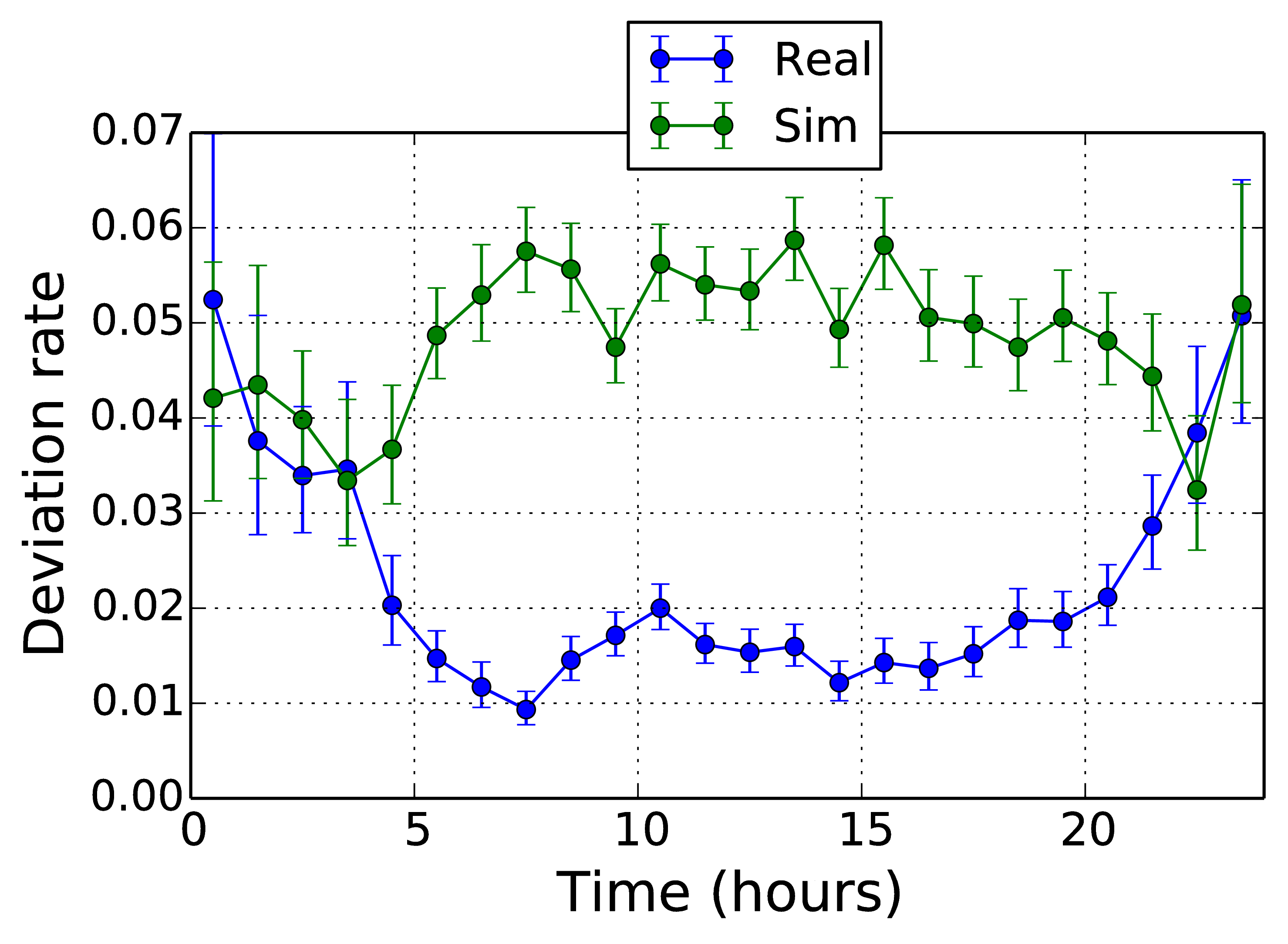} 
                 \caption{Left panel: deviation rates in the ``No Directs'' case. Right panel: deviation rates in the ``No Sector Directs'' case.}\label{fig:mappenocal}
\end{figure}

%%%%%%%%%%%%%%%%%%%%%%%%%%%%%%%%%%%%%%%%%%%%%%%%%%%%%%%%%%%%%%%%%%%%%%%%%%%%%%%%%%%%%%%%%%%%%%%%%%%%
%%%%%%%%%%%%%%%%%%%%%%%%%%%%%%%%%%%%%%%%%%%%%%%%%%%%%%%%%%%%%%%%%%%%%%%%%%%%%%%%%%%%%%%%%%%%%%%%%%%%
\section{Statistical regularities of ABM simulations} \label{results}
%%%%%%%%%%%%%%%%%%%%%%%%%%%%%%%%%%%%%%%%%%%%%%%%%%%%%%%%%%%%%%%%%%%%%%%%%%%%%%%%%%%%%%%%%%%%%%%%%%%%
%%%%%%%%%%%%%%%%%%%%%%%%%%%%%%%%%%%%%%%%%%%%%%%%%%%%%%%%%%%%%%%%%%%%%%%%%%%%%%%%%%%%%%%%%%%%%%%%%%%%

In this section we give some examples of the simulation outputs  of our model obtained with the parameters of the calibration procedure of section \ref{calibration} for the evolution of the planned flight trajectories of the LIRR ACC (Rome, Italy) of the AIRAC 334.

In Fig. \ref{fig:operate} we show the fraction of the different decisions taken by controllers. The three decisions controllers can take are (i) issuing a direct, (ii) re-routing a flight trajectory, and (iii) temporary change the flight level of a trajectory. We label the total number of operations in a given one hour interval as $N_O$. $N_D$ is the number of directs issued by controllers in the time interval. Similarly, $N_R$ is the number of re-routings and $N_F$ is the number of flight level changes.  In Fig. \ref{fig:operate} we show the ratio of directs  $N_D/N_O$ (blue circles), the ratio of re-routings  $N_R/N_O$ (green circles), and the ratio of flight level changes $N_F/N_O$ (red circles).
The error bar is to the 95\% Wilson confidence interval. The ratio of flight level changes (red circles) and the ratio of re-routings (green circles) issued to solve possible conflicts are larger during day-time rather than during night-time. It is worth noting that the number of re-routing is always higher than the number of flight level changes. This is a satisfactory outcome of our model consistent with the feedback we have received from ATM experts. The ratio of directs (blue circles) behaves in the opposite way. This is again expected, given the fact that lower traffic conditions during night allows for the possibility of optimizing trajectories more easily \cite{pap47}. During day-time, the sector workload can be different for different sectors and therefore maximal sector capacity is not reached at the same time for all sectors. This can be an explanation why directs are also issued during day-time.
\begin{figure}[H]
\centering
                 \includegraphics[width=10.0cm]{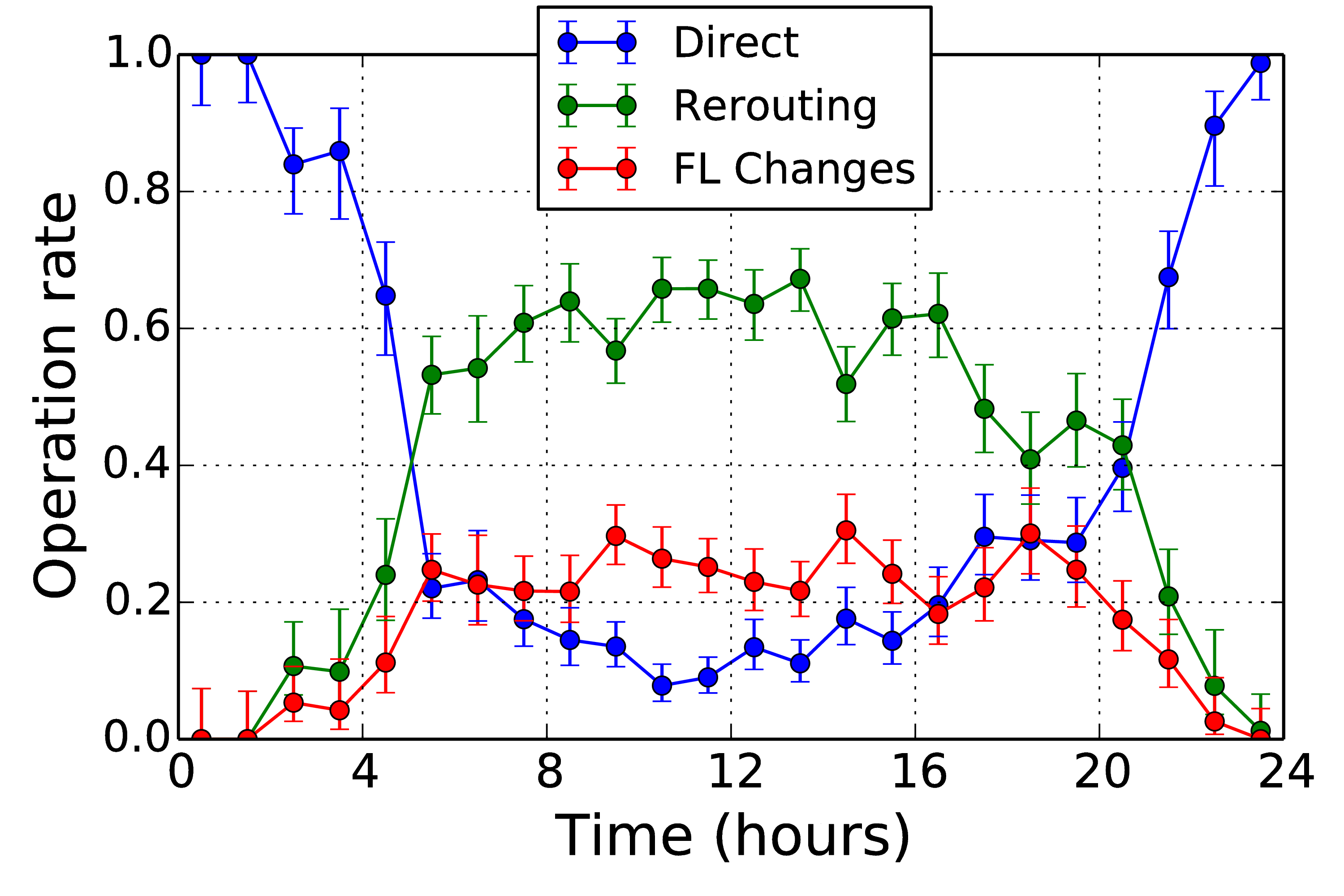}
                 \caption{Controllers' operation rate: the ratio $N_F/N_O$ between the number $N_F$ of flight level changes (red circles) and the total number of operations $N_O$, the ratio $N_R/N_O$ between the number $N_R$ of re-routings (green circles) issued to solve possible conflicts and $N_O$, the ratio $N_D/N_O$ between the number $N_D$ of directs (blue circles) and $N_O$, where $N_O=N_F+N_R+N_D$. The error bar correspond to the 95\% Wilson confidence interval.}\label{fig:operate}
\end{figure}

%%%%%%%%%%%%%%%%%%%%%%%%%%%%%%%%%%%%%%%%%%%%%%%%
\subsection{Conflict resolution in the ABM} \label{DST}
%%%%%%%%%%%%%%%%%%%%%%%%%%%%%%%%%%%%%%%%%%%%%%%%%%%%%%%%%%%%%%%%%%%%%%%%%%%%%%%%%%%%%%%%%%%%%%%%%%%%

In this section we discuss the ability of our model in performing conflict resolution by investigating the distance observed between all pairs of aircrafts flying during a given day.

In Fig. \ref{fig:distance} we show the cumulated distribution of the distance between any pair of aircrafts for a simulation performed for the first day of AIRAC 334. The red curve shows the distribution of the planned trajectories, the blue curve (labeled as simulation I) is the cumulated distribution of the flight trajectories simulated with our model by using the safety threshold of 5 NM. The green curve (labeled as simulation II) is the cumulated distribution of the flight trajectories simulated with our model by using a  safety threshold that increases with the look-ahead, as described in section \ref{deconfmodule}. Specifically, in the second simulation we set $\Delta d_{thr}=0.33$.  

In the figure we highlight as a vertical line the value of 5 NM. It is worth noting that both the blue and the green lines show values that are on the right of the vertical line. This means that our ABM solves all conflicts that were present in the planned flight of the day. The blue line presents values that are quite close to the 5 NM threshold whereas, as expected, the green line has lower values for distances slight above 5NM, thus indicating that aircraft are more separated. 
\begin{figure}[H]
\centering
                 \includegraphics[width=9.0cm]{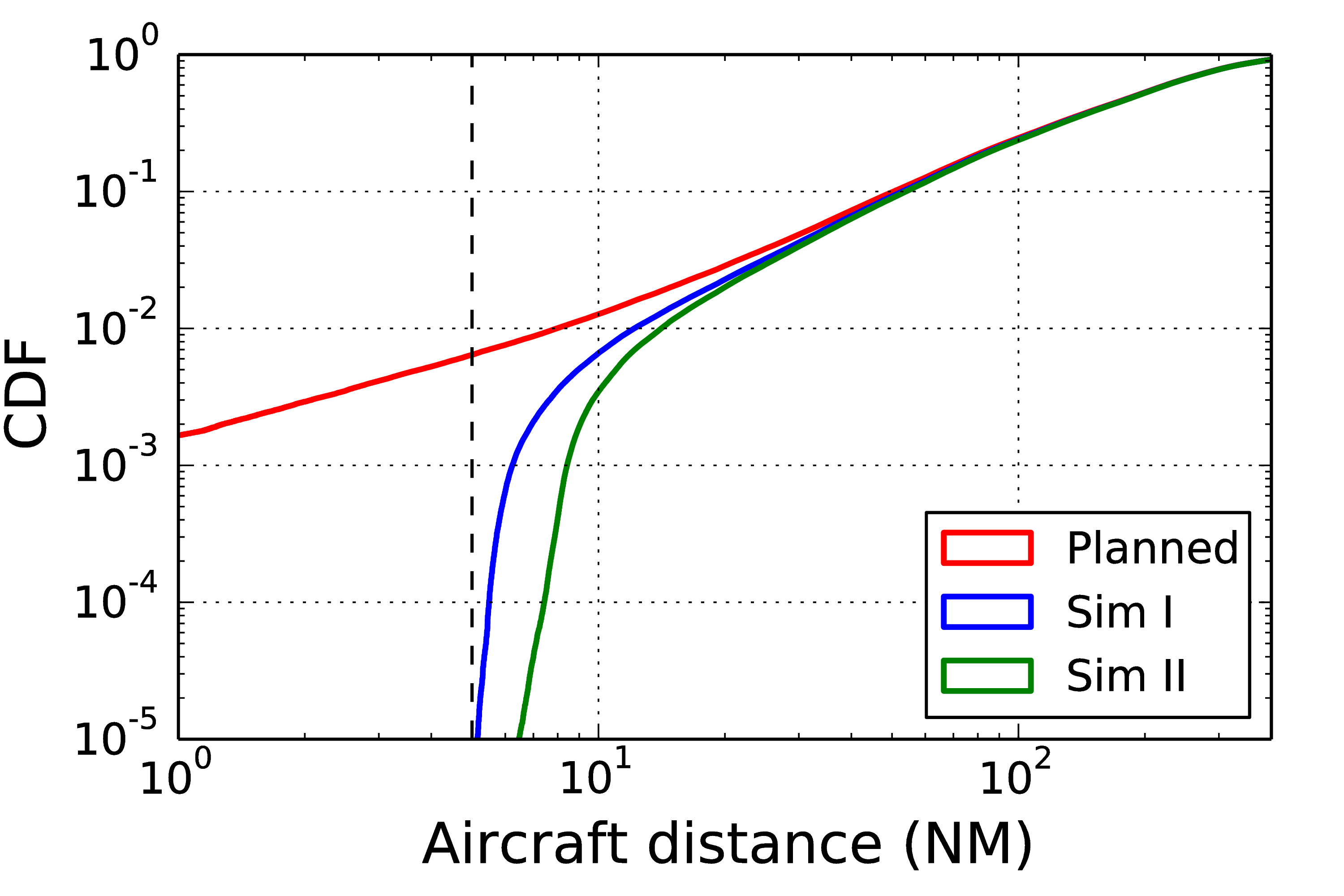} 
                 \caption{Cumulative distribution of the distance between any pair of aircrafts. The red curve shows the cumulative distribution of the planned trajectories, the blue curve (simulation I) is the cumulative distribution of flight trajectories simulated with our model by considering a fixed safety threshold of 5 NM. The green curve (simulation II) is the cumulative distribution of flight trajectories simulated with our model when the safety threshold increases with the look-ahead by setting $\Delta d_{the}=0.33$. The cumulative distribution is obtained by considering all flights of AIRAC 334 and the associated simulations.}
\label{fig:distance}
\end{figure}

The parameter $\Delta d_{thr}$ therefore allows the model to fine tune the probability of observing a pair of aircrafts with a given minimal distance in a given day. As also recalled in Table \ref{modpar}, $\Delta d_{thr}$ is a parameter that might in principle reflect the ATCOs behavior when managing traffic with a large look-ahead.  In fact, a large $\Delta d_{thr}$ would indicate that human controllers tend to be overly safe when managing trajectories with a large look-ahead and tend to separate aircraft pairs more than it is needed. The model shows that this might end up in having aircraft separated more the 5 NM and therefore in a non-optimal usage of the available airspace that in turn leads to a reduction of the maximal sector capacity. On the other hand, a small $\Delta d_{thr}$ would indicate that human controllers are rather confident about their procedures even for aircraft that are far away. In this case our simulations indicate that all available airspace is used which might lead to an optimal assessment of sector capacity.

%%%%%%%%%%%%%%%%%%%%%%%%%%%%%%%%%%%%%%%%%%%%%%%%%%%%%%%%%%%%%%%%%%%%%%%%%%%%%%%%%%%%%%%%%%%%%%%%%%%%
\subsection{Spatial heterogeneity of the operations} \label{operation hubs}
%%%%%%%%%%%%%%%%%%%%%%%%%%%%%%%%%%%%%%%%%%%%%%%%%%%%%%%%%%%%%%%%%%%%%%%%%%%%%%%%%%%%%%%%%%%%%%%%%%%%

In Fig. \ref{fig:mappenavigation points} we show the map of navigation points with the information about the type of operations controllers do in their neighboring. In the left panel we show re-routings.  In the central panel we show flight level changes, while in the right panel we show directs. In all panels the size of of circles is proportional to the number of operations performed. All values refer to the 334 AIRAC. Interestingly, the navigation points with the highest number of re-routings are aligned along the route between Milan and Rome, which shows the highest traffic levels, as indicated in Fig. \ref{fig:DistributionDev}. On the other hand the highest number of directs is issued either in central Italy (most probably in proximity of Fiumicino airport) or in the Thyrrenian Sea, between Naples and Sicily, where traffic levels are less pronounced than in the northern region, as indicated in Fig. \ref{fig:DistributionDev}. The location of flight level change operations highlights those navigation points where controllers have difficulties in solving conflicts and use flight level change as the last resort for conflict solution. 
\begin{figure}[H]
\centering
                 \includegraphics[width=5.0cm]{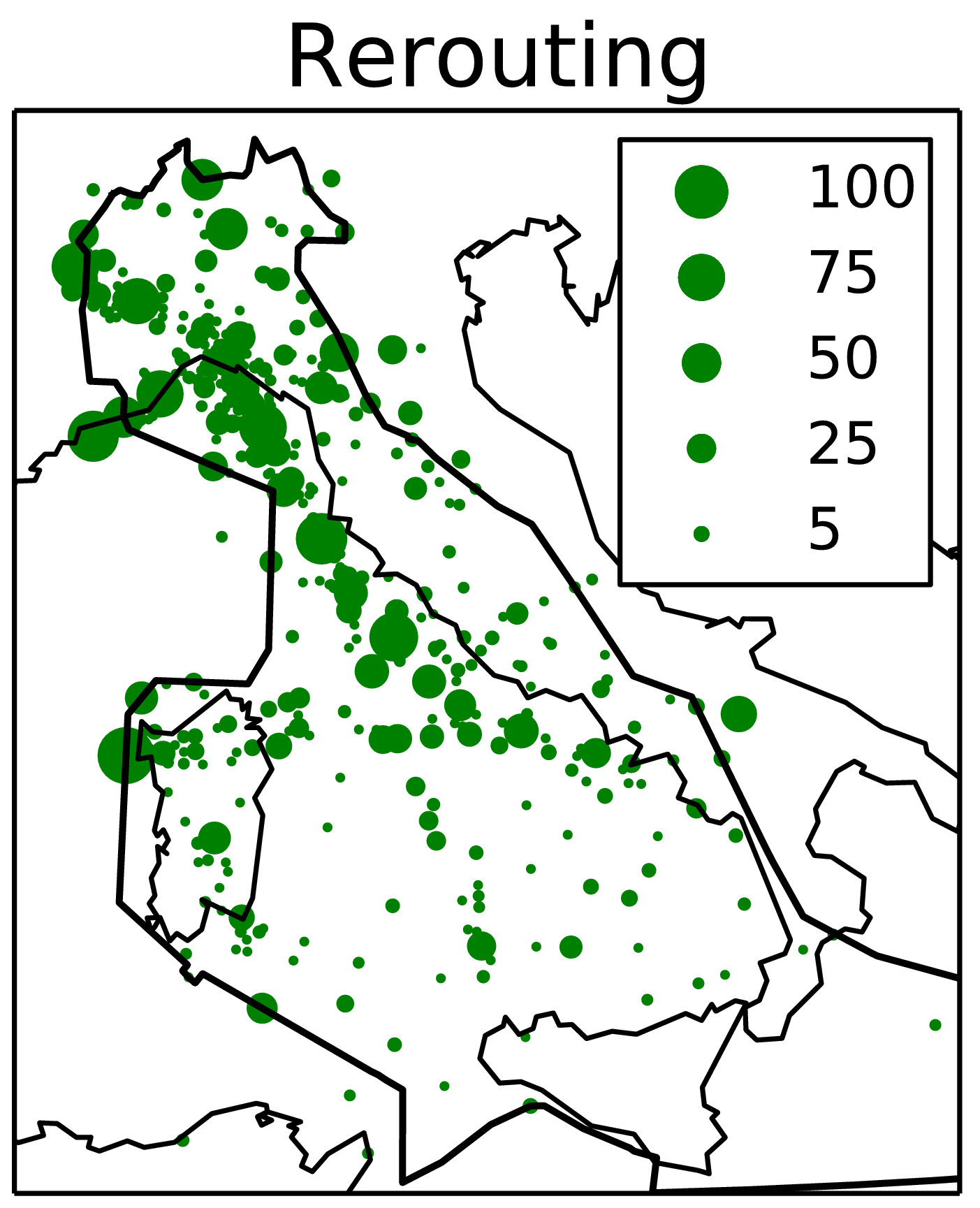} 
                 \includegraphics[width=5.0cm]{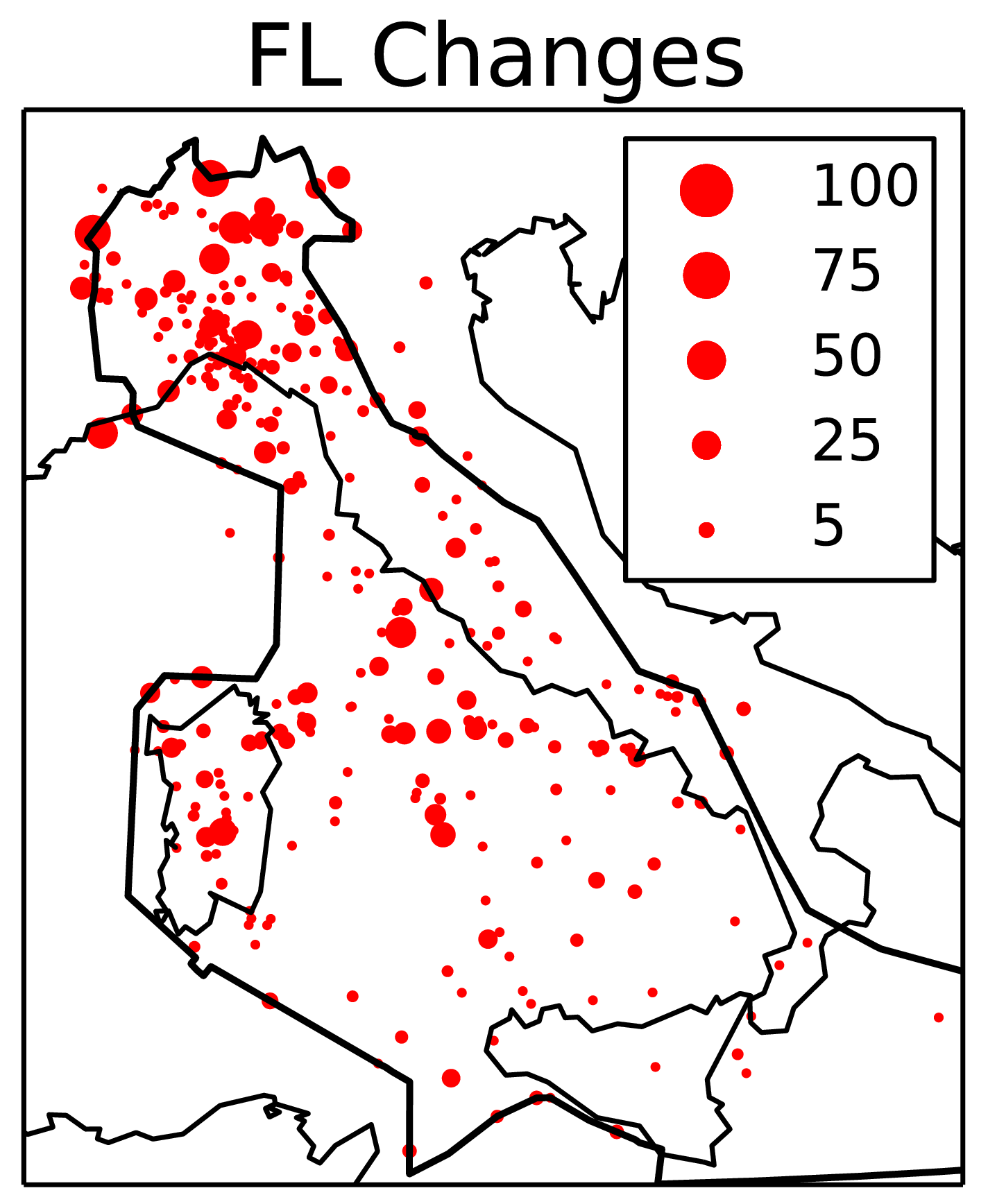}
                 \includegraphics[width=5.0cm]{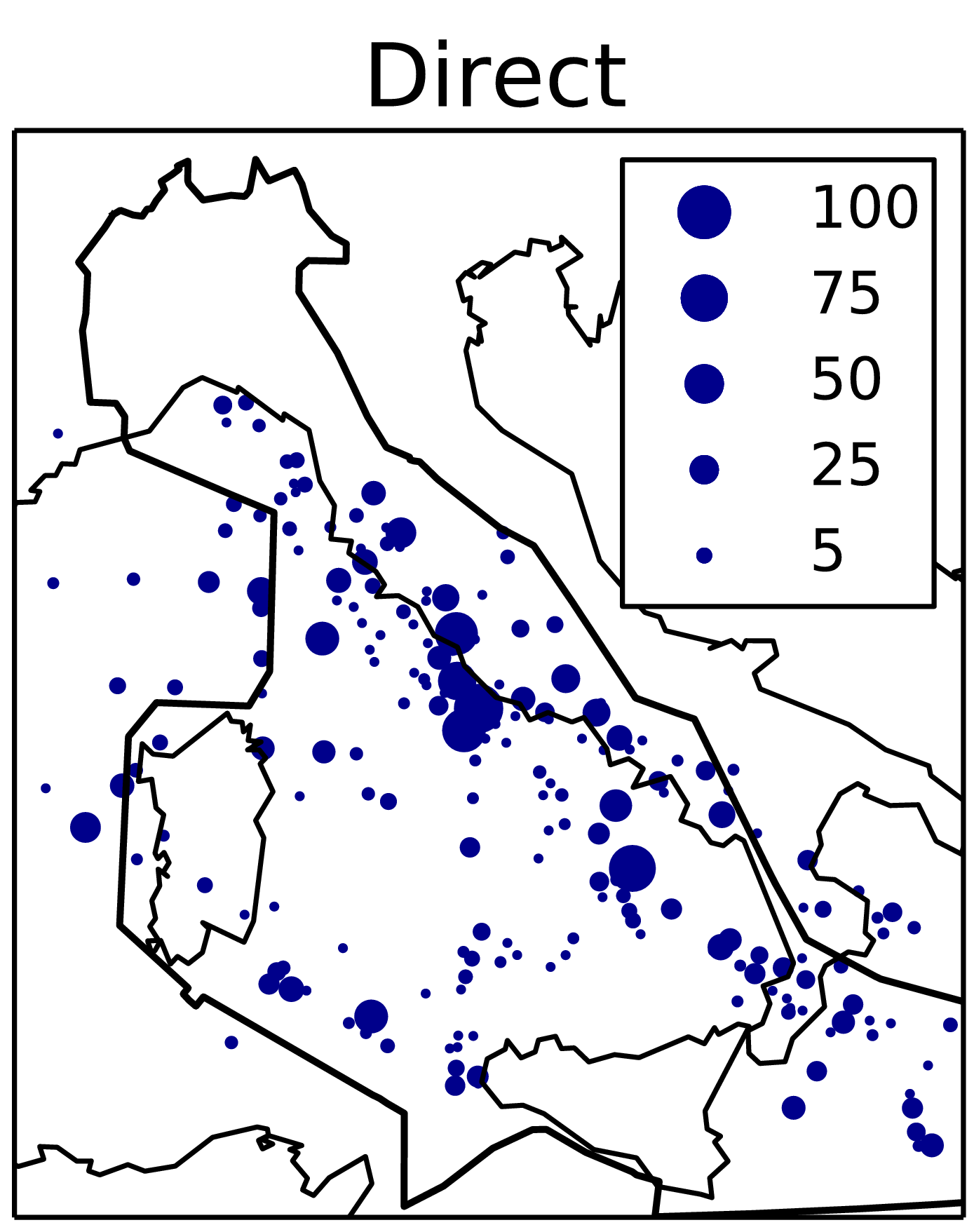} 
                 \caption{Map of navigation points with the information about the type of operations controllers have to do in their neighboring. Left panel (green): the size of each circle is proportional to the number of re-routings. Central panel (red): the size of each circle is proportional to the number of  flight level changes. Right panel (blue): the size of each circle is proportional to the number of directs}\label{fig:mappenavigation points}
\end{figure}

A similar result also holds for operations performed by real ATCOs. Indeed, the ATCO operations do not uniformly affect the flux of aircraft in the airspace. Rather, ATCOs typically concentrate their operations on specific segments of flight trajectories (i.e. on the path joining two neighboring navigation points). This is clearly shown by the results summarized in Fig.\ref{fig:DistributionDev} where we show the distribution of the difference $M=M_{pp}-M_{pr}$ between the number of planned flights that should have passed through a certain trajectory segment $M_{pp}$ and the number of these flights that actually passed through that  trajectory segment $M_{pr}$. The blue line shows empirical data, while the green line refers to data obtained through numerical simulations of our ABM. The red line refers to a random allocation of $M$ values the missed flight in each trajectory segment. This random allocation preserves $(i)$ the planned number of flights in each trajectory segment $M_{pp}$ and $(ii)$ the sum $\sum_{link} M$ for the whole ACC. Such random sampling therefore preserves the planned heterogeneity of the system as well as the global number of operations done by the controllers. 

Two comments are in order. On one hand, one can notice that the ABM well reproduces the empirical observations. On the other hand, it is worth noticing that these two distributions show tails that are fatter than those of the distribution obtained with the random sampling. This indicates that there are trajectory segments where the number of operations done by the controllers is higher that what should be expected by the random null model. This clearly suggests that ATCO operations tend to be focused on specific regions of the ACC. The comparison with such simple random null model therefore allow us to highlight the presence of specific regions in the airspace that cannot be explained just with the heterogeneity of the flux of aircrafts: it is therefore a genuine effect produced by the ATCOs and it is quite well reproduced by the ABM.
\begin{figure}[H]
\centering
 	\includegraphics[width=7.5cm]{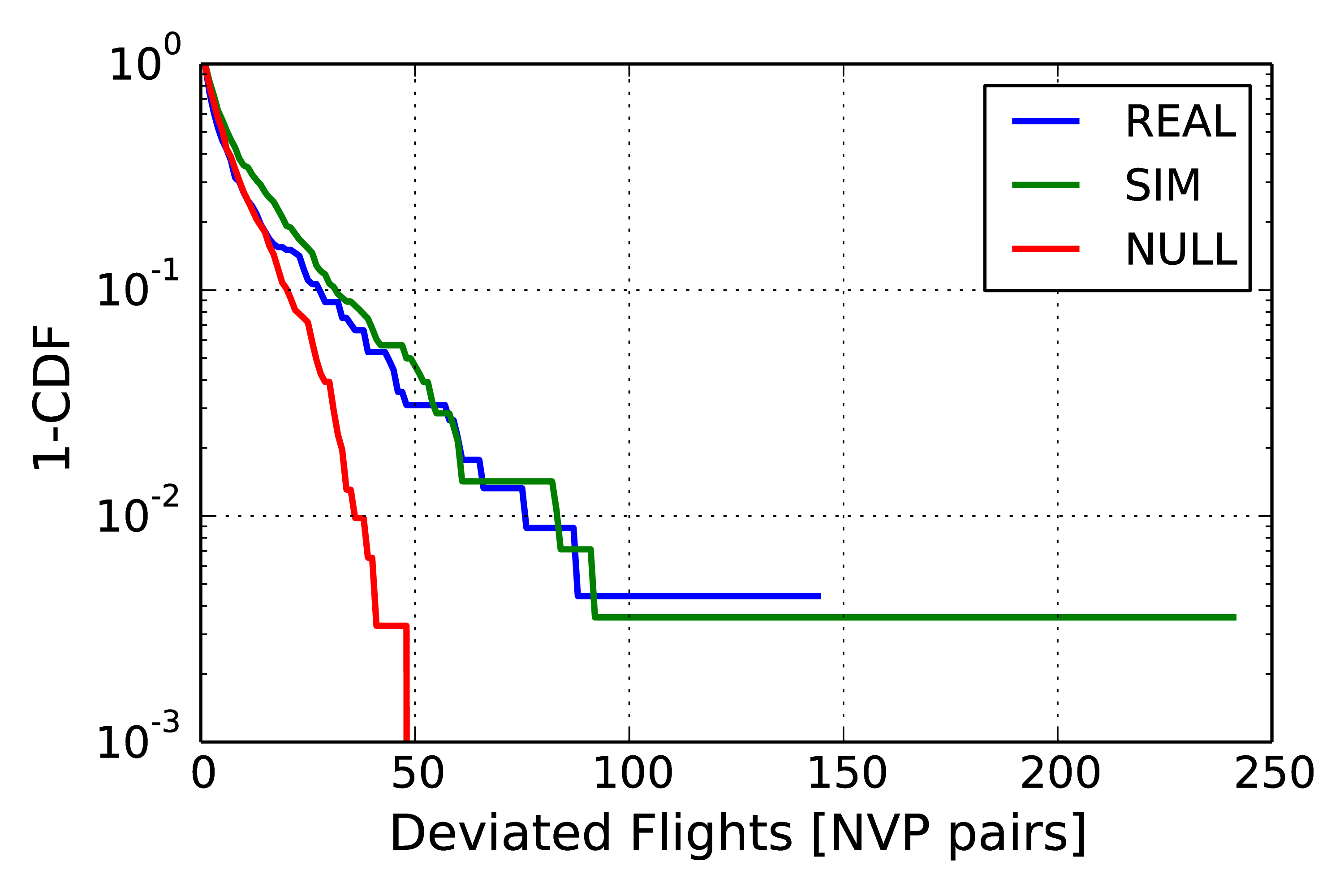} 
        \caption{Complementary cumulative distribution of the difference $M=M_{pp}-M_{pr}$ between the number of planned flights that should have passed through a certain trajectory segment $M_{pp}$ and the number of these flights that actually passed through that  trajectory segment $M_{pr}$. The blue line refer to empirical data, while the green curve refer to data obtained through numerical simulations of our ABM. The red curve refer to data obtained by performing a random sampling of the missed flight in each trajectory segment.}\label{fig:DistributionDev}
\end{figure}

However, although the ABM well reproduces the existence of regional heterogeneity,  it is worth emphasizing that there are airspace regions where the ABM and human ATCOs manage traffic in a different way. In Fig. \ref{fig:diff} we show the difference M  in a specific region of the ACC located close to Genoa and characterized by high traffic conditions. The left panel refers to the empirical case while the right panel refers to numerical simulations performed with our ABM. The difference M is here shown through a color scale reported on the right of each panel. One can see that there are trajectory segments where ATCOs do not modify planned trajectories (lighter colors) that are instead quite heavily affected by the ABM (darker colors) and viceversa.
\begin{figure}[H]
\centering
                 \includegraphics[width=8cm]{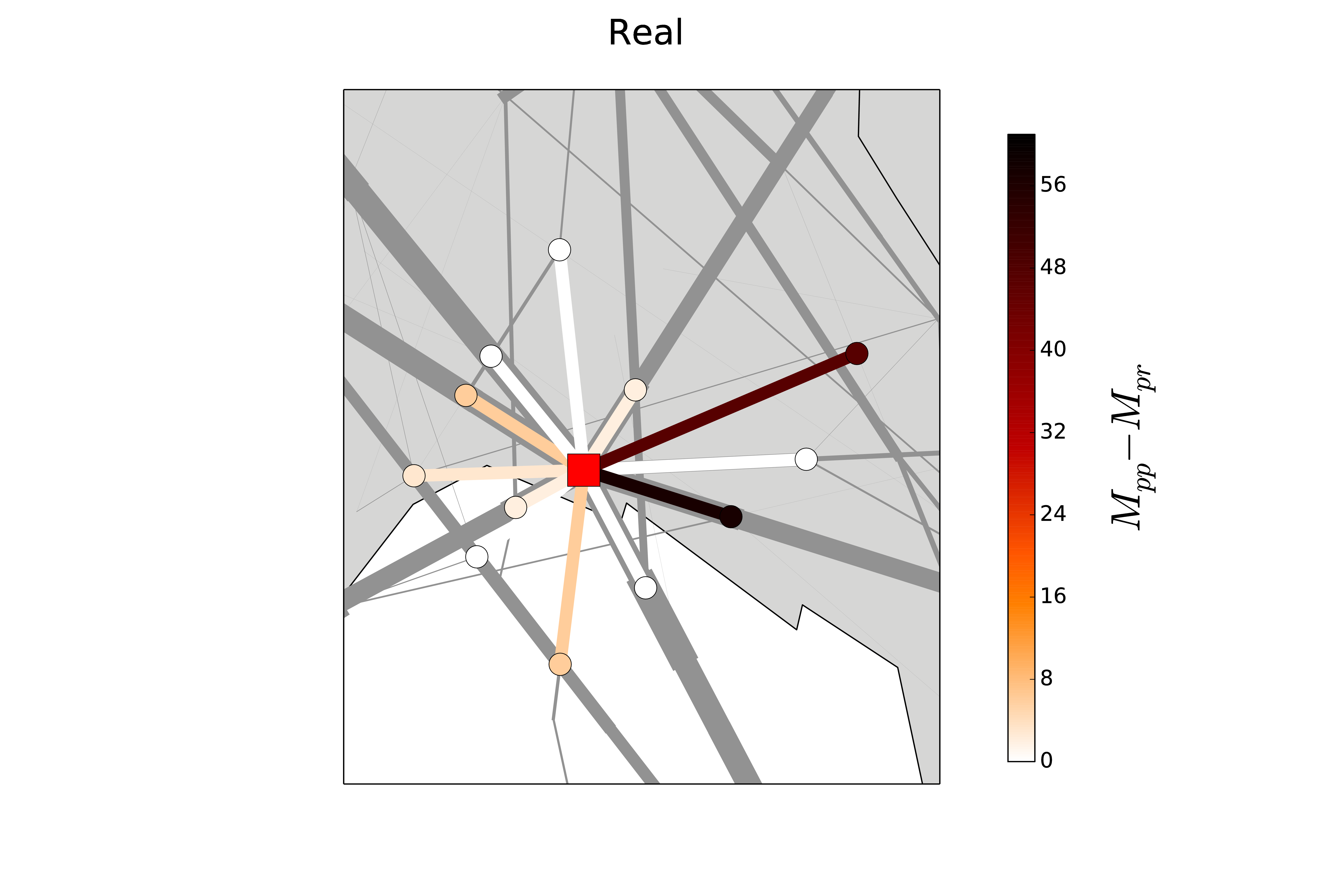}
                 \includegraphics[width=8cm]{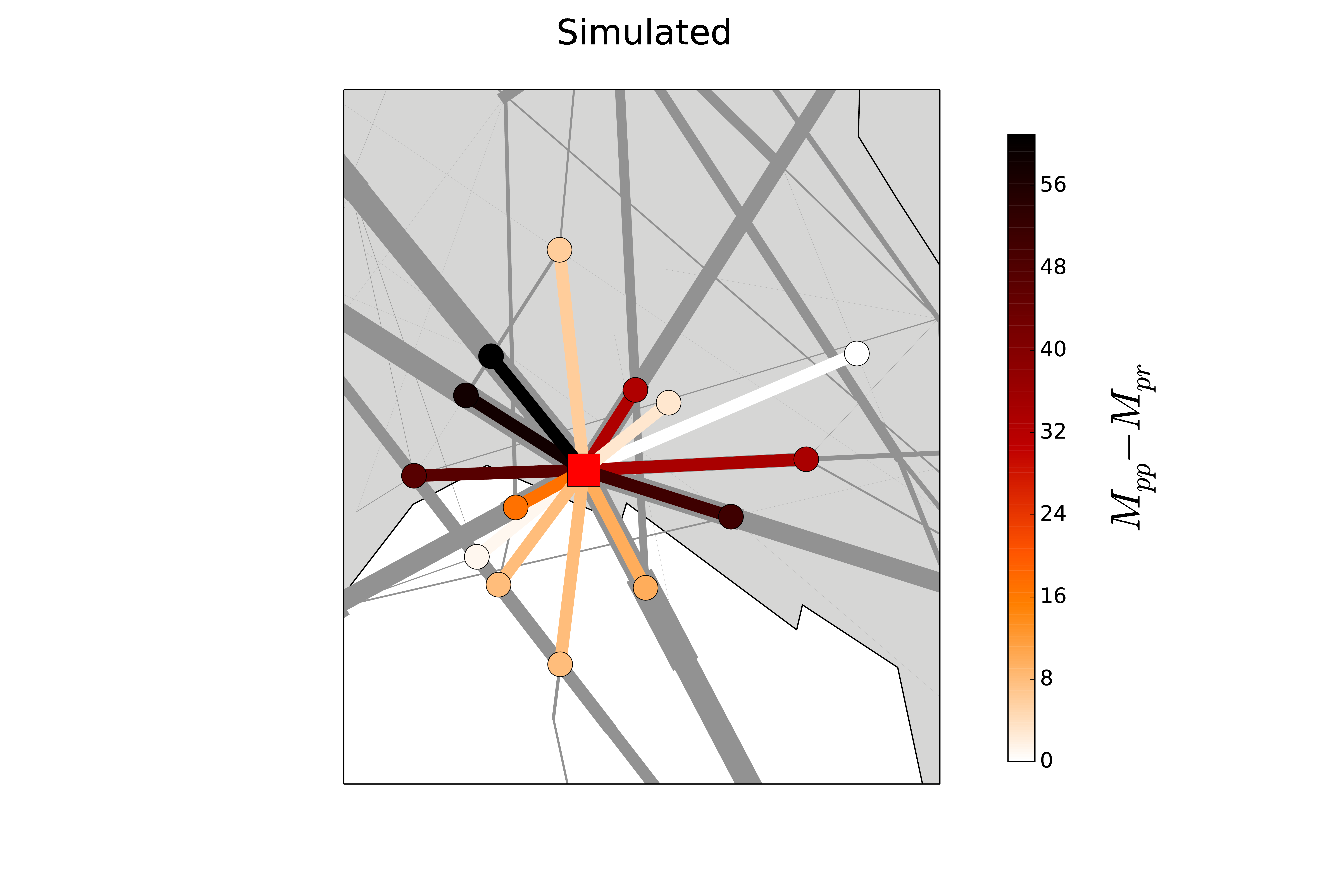} 
                 \caption{Difference M  in a specific region of the ACC located close to Genoa, Italy. The left panel refers to the empirical case while the right panel refers to numerical simulations performed with our ABM. The difference M is here shown through the color-code reported on the right of each panel.}\label{fig:diff}
\end{figure}

In fact, this should not be surprising given the fact that ATCOs have to deal with tactical conditions (weather events, aircraft problems, ....) that our ABM does not take into account. Moreover, this different behavior might also be due to the fact that human controllers tend to be overly safe and therefore have a conservative style in managing the aircraft trajectories.

%%%%%%%%%%%%%%%%%%%%%%%%%%%%%%%%%%%%%%%%%%%%%%%%%%%%%%%%%%%%%%%%%%%%%%%%%%%%%%%%%%%%%%%%%%%%%%%%%%%%
%%%%%%%%%%%%%%%%%%%%%%%%%%%%%%%%%%%%%%%%%%%%%%%%%%%%%%%%%%%%%%%%%%%%%%%%%%%%%%%%%%%%%%%%%%%%%%%%%%%%
\section{Dependence of directs and conflict resolution rates from model parameters} \label{directsafety}
%%%%%%%%%%%%%%%%%%%%%%%%%%%%%%%%%%%%%%%%%%%%%%%%%%%%%%%%%%%%%%%%%%%%%%%%%%%%%%%%%%%%%%%%%%%%%%%%%%%%
%%%%%%%%%%%%%%%%%%%%%%%%%%%%%%%%%%%%%%%%%%%%%%%%%%%%%%%%%%%%%%%%%%%%%%%%%%%%%%%%%%%%%%%%%%%%%%%%%%%%

Finally, we report on how our model performs under parameters different from the ones chosen for calibration. Specifically, we evaluate the performances of our model with respect to model decisions concerning directs and conflict resolutions as a function of procedures followed by air traffic controllers and air traffic conditions of sectors.

Results of our investigation are summarized in Fig. \ref{fig:safetydirects}. In the left panel of Fig. \ref{fig:safetydirects} we show the number of actions that the controllers perform  in order to solve conflicts, i.e. re-routings and flight level changes, as a function of the number of directs for the five values of $\Delta t$ shown in the legend. Each point in the plot corresponds to the result of a simulation of the ABM performed with a pair $(p_d, x_c)$ of parameters selected in the range $p_d \in [0.03,0.5]$
(with step of 0.01567), $x_c \in [0.34,1.5]$ (with step of 0.03867). The figure suggests the existence of a linear negative relation between the number of operations needed to solve conflicts and the number of directs, thus indicating that the number of unsolved conflicts decreases when the number of directs issued increases. 

These results refer to ATCOs able to do a perfect forecast within the look-ahead used when directs are issued $\Delta t_d$. In reality, many unexpected factors can contribute to make uncertain a forecast. Uncertainty can result for example from a flight entering the airspace within $\Delta t_d$ unexpectedly or a weather event, or some errors in the forecast of aircraft positions. We evaluate the performance of our ABM model with respect to this type of uncertainty by performing a series of simulations in the presence of a source of noise. Specifically, the source of noise is introduced in the velocity of aircrafts. In the right panel of Fig. \ref{fig:safetydirects} we show the results of a numerical simulation obtained by introducing noise in the velocity estimation of the aircraft. The parameter used is $l_{\epsilon} = 0.1$ which is a quite large value. This produces the effect of increasing the number of needed conflict resolutions especially for simulations with a high value of the look-ahead. 
\begin{figure}[H]
\centering
                 \includegraphics[width=8.5cm]{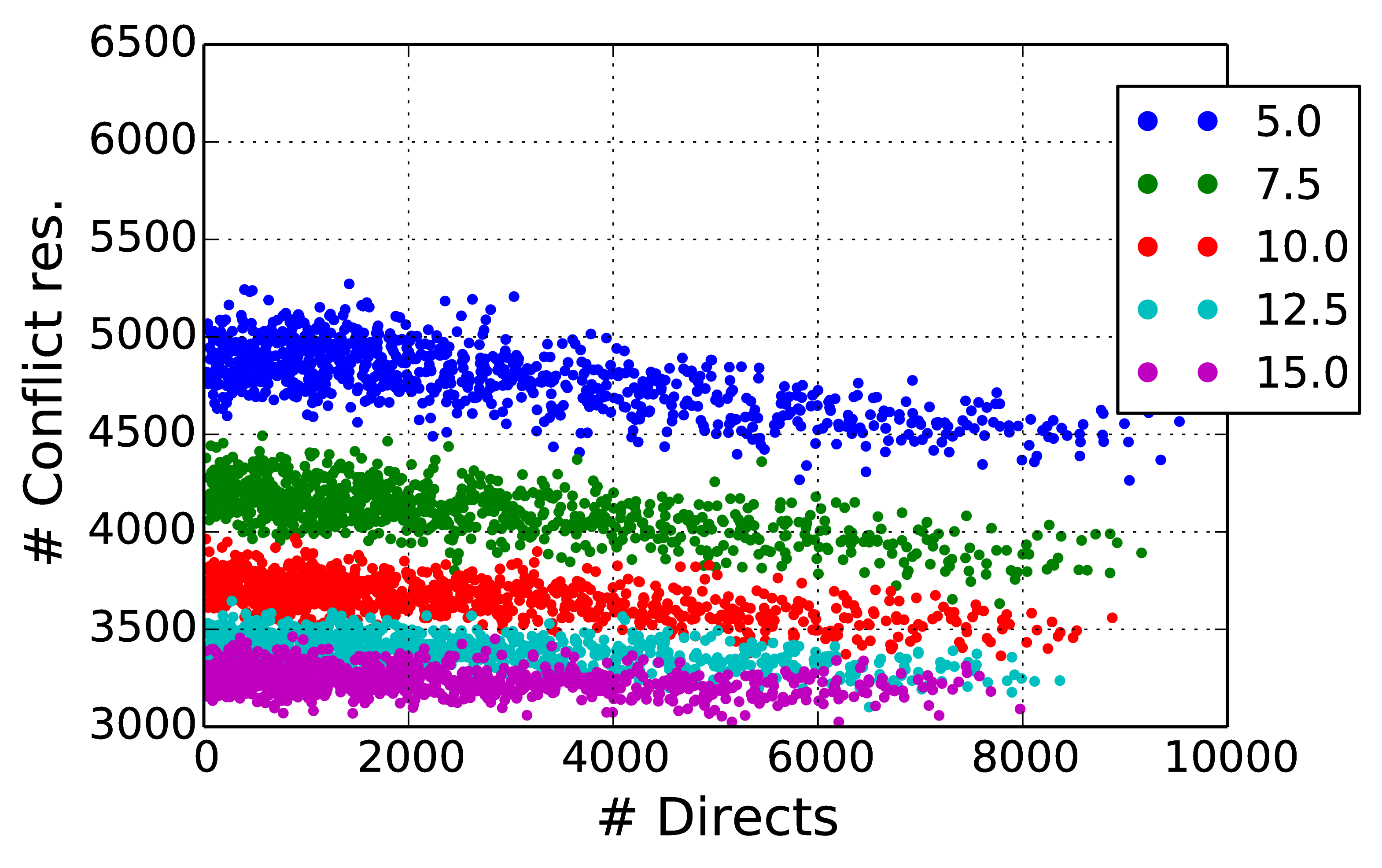} 
                 \includegraphics[width=8.5cm]{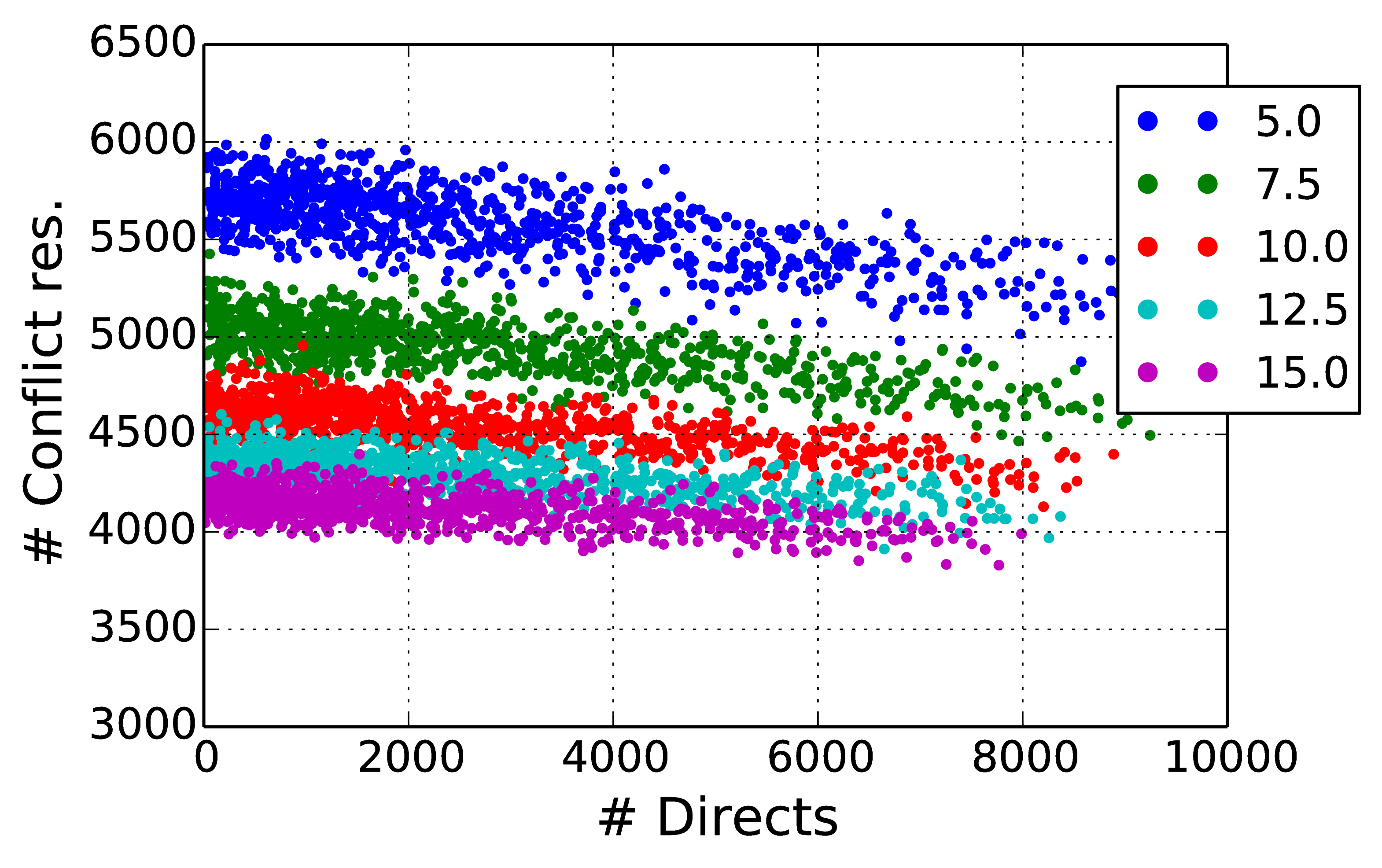} 
                 \caption{Number of conflict resolutions as a function of the number of issued directs. Each point in the plot corresponds to the result of a simulation of the ABM performed with a pair $(p_d, x_c)$ of parameters selected in the range $p_d \in [0.03,0.5]$
(with step of 0.01567), $x_c \in [0.34,1.5]$ (with step of 0.03867). Different colors refers to different values of the look-ahed $\Delta t_l$.  In all simulations $\Delta t_d=15$ min.The simulations on the left panel are made with a perfect forecast. The simulations on the left panel are done by introducing a noise in the velocity of aircrafts ($l_\epsilon=0.1$). This noise strongly affects the reliability of forecast of flight trajectories. }\label{fig:safetydirects}
\end{figure}

In Table \ref{fit} we report the result of a linear fitting procedure on the five sets of simulations obtained for different values of $\Delta t_l$ and shown in Fig. \ref{fig:safetydirects} as points of different colors. The upper part of the table refers to simulations with perfect forecast whereas the lower part refers to simulations in the presence of noise. The $p-value$ reported in a column is the two-sided $p-value$ of the null hypothesis that the slope of the linear relationship is zero.  Indeed, the low $p-values$ observed support the existence of a linear relationship between directs and conflict resolution events, although the slope value can be quite small in all considered cases. In fact, the correlation values reported in the fourth column are indicating a statistically robust negative relationship between directs and conflict resolution events. 
%Once again, this results emphasize the need of a correct calibration procedure in order to have a realistic representation of the ATM system. 
\begin{table}[H]
\centering
\begin{tabular}{|c|c||c|c|c|c|c|}
\hline
{\bf{lookahed (min)}}  & {\bf{noise}} & {\bf{slope}}  & {\bf{intercept}}  & {\bf{correlation coef.}}  & {\bf{p-value}}  & {\bf{std err}}\\
\hline
5.00 & 0.0	& -0.052   & 4939   & -0.691 &  10$^{-128}$  & 0.002  \\
7.50 & 0.0 	& -0.040   & 4215   & -0.653 & 10$^{-110}$   & 0.002  \\
10.0 & 0.0	& -0.028   & 3734   & -0.571 & 10$^{-79}$   	& 0.001 \\
12.5 & 0.0	& -0.021   & 3444  & -0.522 &  10$^{-64}$   & 0.001 \\
15.0 &	0.0	& -0.013   & 3273   & -0.342 &  10$^{-26}$   & 0.001 \\
\hline
5.00 & 0.1	& -0.060		& 5749   & -0.719 &  10$^{-144}$  & 0.002  \\
7.50 & 0.1 	& -0.046   	& 5073   & -0.694 & 10$^{-130}$   & 0.002  \\
10.0 & 0.1	& -0.039   	& 4650   & -0.685 & 10$^{-125}$   & 0.001 \\
12.5 & 0.1	& -0.030   	& 4370   & -0.582 &  10$^{-82}$   & 0.001 \\
15.0 & 0.1	& -0.027   	& 4186   & -0.555 &  10$^{-74}$   & 0.001 \\
\hline

\hline
\end{tabular}
\caption{Summary statistics of the result of a linear fitting procedure on the five sets of simulations obtained with different values of $\Delta t_l$. Other parameters are changed as described in the text.  The upper part of the table refers to simulations with perfect forecast whereas the lower part refers to simulations in the presence of noise. The simulations in the presence of noise are obtained by setting $l_\epsilon=0.1$.}  \label{fit}
\end{table}

It is worth noting that slopes observed in the presence of noise are systematically higher in absolute value that in the case of perfect forecast. This seems to suggest that also in the presence of enhanced uncertainty issuing directs reduces the number of conflicts to be resolved.

%%%%%%%%%%%%%%%%%%%%%%%%%%%%%%%%%%%%%%%%%%%%%%%%%%%%%%%%%%%%%%%%%%%%%%%%%%%%%%%%%%%%%%%%%%%%%%%%%%%%
%%%%%%%%%%%%%%%%%%%%%%%%%%%%%%%%%%%%%%%%%%%%%%%%%%%%%%%%%%%%%%%%%%%%%%%%%%%%%%%%%%%%%%%%%%%%%%%%%%%%
\section{Conclusions} \label{concl}
%%%%%%%%%%%%%%%%%%%%%%%%%%%%%%%%%%%%%%%%%%%%%%%%%%%%%%%%%%%%%%%%%%%%%%%%%%%%%%%%%%%%%%%%%%%%%%%%%%%%
%%%%%%%%%%%%%%%%%%%%%%%%%%%%%%%%%%%%%%%%%%%%%%%%%%%%%%%%%%%%%%%%%%%%%%%%%%%%%%%%%%%%%%%%%%%%%%%%%%%%

In this work we have presented an agent-based model of the ATM system that aims at modeling the interactions between aircrafts and ATC controllers at a tactical level. We have presented in detail the different modules of the model whose core is given by the conflict detection and resolution module of section \ref{deconfmodule} and by the directs module of section \ref{directsD}.

In section \ref{calibration} we have given an example of the calibration of our model 
done in order to obtain simulations describing the statistical regularities about the rate of flight trajectory deviations observed in empirical data.

In section \ref{results} we have reported results obtained with our model.
First, we explicitly show that the calibrated model is able to reproduce the existence of regional localization of ATM operations, i.e. the fact that ATCO operations tend to be focused on specific points of the ACC. Finally, we have shown scenario simulations results about the relationship between directs and  conflict resolution events conditioned to model parameters. 

Our model can be used to give useful insights about the functioning of the ATM system.  We are aware that our model is very basic. For example, our basic agent based model does not implement any learning mechanism as done for example in other models \cite{agogino} or specific fitness measures besides the fact that in the conflict resolution module we follow must consider the shortest trajectory amongst the possible ones. Furthermore, the model implements a local resolution of conflicts. The way our model solves conflicts is fast from a computational point of view but provides solutions that are not optimized at a global level, thus making it necessary to check trajectories several times as long as an aircraft travels across the ACC. We are fully aware of this limitation of our model. We implemented such a solution because we wanted to develop an ABM mimicking the way air traffic controllers work in reality. 

Indeed, we  believe that such solution might be quite effective in the SESAR scenario simulations. In fact, we might simulate a scenario where controllers have a role less preeminent than in the current scenario and some basic conflict-resolution actions are left to the single aircraft. In this respect, our model might mimic a scenario where pilots,  that clearly have not a global vision of the system, endowed with a set of policy rules assigned by their airlines, will perform an {\em{active}} conflict resolution at a tactical level, thus realizing a sort of self-organization amongst aircraft.  Along similar lines, other possible ways for further research starting from the present model regard the possibility of augmenting our model capabilities by implementing learning and self-adaptation mechanisms as well as some level of intelligence for the agents.

%%%%%%%%%%%%%%%%%%%%%%%%%%%%%%%%%%%%%%%%%%%%%%%%%%%%%%%%%%%%%%%%%%%%%%%%%%%%%%%%%%%%%%%%%%%%%%%%%%%%
%%%%%%%%%%%%%%%%%%%%%%%%%%%%%%%%%%%%%%%%%%%%%%%%%%%%%%%%%%%%%%%%%%%%%%%%%%%%%%%%%%%%%%%%%%%%%%%%%%%%
\acknowledgments
This work was co-financed by EUROCONTROL on behalf of the SESAR Joint Undertaking in the context of SESAR Work Package E - ELSA research project. We thank the Deep Blue Srl team for providing us useful insights about the ATM system. We also thank all past participants to the ELSA project for  interesting discussions.

%%%%%%%%%%%%%%%%%%%%%%%%%%%%%%%%%%%%%%%%%%%%%%%%%%%%%%%%%%%%%%%%%%%%%%%%%%%%%%%%%%%%%%%%%%%%%%%%%%%%
%%%%%%%%%%%%%%%%%%%%%%%%%%%%%%%%%%%%%%%%%%%%%%%%%%%%%%%%%%%%%%%%%%%%%%%%%%%%%%%%%%%%%%%%%%%%%%%%%%%%
\section*{ADDITIONAL INFORMATION}
The authors have no competing interests as defined by Nature Publishing Group, or other interests that might be perceived to influence the results and/or discussion reported in this paper.

%%%%%%%%%%%%%%%%%%%%%%%%%%%%%%%%%%%%%%%%%%%%%%%%%%%%%%%%%%%%%%%%%%%%%%%%%%%%%%%%%%%%%%%%%%%%%%%%%%%%
%%%%%%%%%%%%%%%%%%%%%%%%%%%%%%%%%%%%%%%%%%%%%%%%%%%%%%%%%%%%%%%%%%%%%%%%%%%%%%%%%%%%%%%%%%%%%%%%%%%%
\section*{Author Contribution}

CB, SM, RNM conceived the study. CB implemented the code and performed numerical simulations. CB, SM, RNM analyzed the data. CB, SM, RNM wrote the manuscript. All authors reviewed the manuscript.

%%%%%%%%%%%%%%%%%%%%%%%%%%%%%%%%%%%%%%%%%%%%%%%%%%%%%%%%%%%%%%%%%%%%%%%%%%%%%%%%%%%%%%%%%%%%%%%%%%%%
%%%%%%%%%%%%%%%%%%%%%%%%%%%%%%%%%%%%%%%%%%%%%%%%%%%%%%%%%%%%%%%%%%%%%%%%%%%%%%%%%%%%%%%%%%%%%%%%%%%%


\begin{thebibliography}{}
%%%%%%%%%%%%%%%%%%%%%%%%%%%%%%%%%%%%%%%%%%%%%%%%%%%%%%%%%%%%%%%%%%%%%%%%%%%%%%%%%%%%%%%%%%%%%%%%%%%%
%%%%%%%%%%%%%%%%%%%%%%%%%%%%%%%%%%%%%%%%%%%%%%%%%%%%%%%%%%%%%%%%%%%%%%%%%%%%%%%%%%%%%%%%%%%%%%%%%%%%

\bibitem{sesar2007}
SESAR (2007), ``SESAR Concept of Operations'' {\tt{https://www.eurocontrol.int/sites/default/files/field\_tabs/content}}\\{\tt{/documents/sesar/20070717-sesar-conops.eps}};
SESAR (2012), ``SESAR Concept of Operations Step 1'' {\tt{http://www.sesarju.eu/sites/default/files/documents/highlight/SESAR\_ConOps\_Document\_Step\_1.eps}}.

\bibitem{comreg}
EU Commission (2010)
``Commission regulation (EU) no 691/2010''.

\bibitem{eurocontrol2005}
EUROCONTROL (2005), ``Final report on european commissions mandate to support the  establishment of functional airspace blocks (fabs)''.

\bibitem{CWPP} 
Cook A.C.,  Rivas D. Eds. (2016)
``Complexity Science in Air Traffic Management''
(Routledge, England, 2016)

%\bibitem{elsa} 
%{\tt{http://complexworld.eu/wiki/ELSA}}

%\bibitem{chen}
%B. Chen, and H. H. Cheng. "A review of the applications of agent technology in traffic and transportation systems." Intelligent Transportation Systems, IEEE Transactions on 11.2 (2010): 485-497.

%\bibitem{schelling}
%T. C. Schelling, 
%Dynamic models of segregation, 
%Journal of mathematical sociology 1 (2), 143-186, (1971). 

%\bibitem{boorman}
%S. A. Boorman, 
%A combinatiorial optimization model for transmission of job information through contact networks, 
%The Bell journal of economics 216-249, (1975)

\bibitem{heat}
Heath B., Hill R., Ciarallo F. (2009) 
 ``A survey of agent-based modeling practices (January 1998 to July 2008)'', 
Journal of Artificial Societies and Social Simulation, {\bf{12}} (4), 9-xx. 

\bibitem{windrum}
Windrum P., Fagiolo G., Moneta A. (2007) 
 ``Empirical validation of agent-based models: Alternatives and prospects'', 
Journal of Artificial Societies and Social Simulation, {\bf{10}} (2), 8-xx.

\bibitem{kuchar}
Kuchar J. K., Yang L. C. (2000)
``A Review of Conflict Detection and Resolution Modeling Methods'', 
IEEE Transactions on Intelligent Transportation Systems, {\bf{1}} (4), 179-189.

\bibitem{agogino}
Agogino A. K., Tumer K. (2012) 
``A multiagent approach to managing air traffic flow'',
Auton Agent Multi-Agent Syst, {\bf{24}}, 1-25.

\bibitem{shah}
Shah A. P., Prichett A. R., Feigh K. M., Kalaver S. A. (2005) 
``Analyzing air traffic management systems using agent based modeling and simulation'', 
Proceedings of the 6th USA/Europe Seminar on Air Traffic Management Research and Development. (Baltimore, 27-30 June 2005, USA).

\bibitem{durand}
Durand N. , Alliot J.-M. (1997)  
``Optimal Resolution of En-Route Conflicts'',  
Proceedings of the 1rst USA/Europe Seminar. (Saclay, 17-19 June 1997, France).

\bibitem{bilimoria1}
Bilimoria K.D.  (2000)  
``A geometric optimization approach to aircraft conflict resolution'',  
Proceedings of the AIAA guidance, navigation, and control conference and exhibit. (Reston, 14-17 August 2000, USA).

\bibitem{eby}
Eby M. S., Kelly W. E. (1999)  
``Free Flight Separation Assurance Using Distributed Algorithms'', 
Proceedings of the IEEE Aerospace Conference, Vol. 2, pp. 429-441. (Snowmass at Aspen, 06-13 March 1999, USA).

%\bibitem{bilimoria2}
%K.D. Bilimoria, B. Sridhar, 
%FACET: Future ATM Concepts Evaluation Tool, 3rd USA/Europe Air Traffic Management R\&D Seminar, 
%Napoli, Italy, 13-16 June 2000

%\bibitem{stream}
%S. Ruiz, M. A. Piera, J. Nosedal, A. Ranieri, 
%Strategic de-confliction in the presence of a large number of 4D trajectories using a causal modeling approach, 
%Transportation Research Part C: Emerging Technologies {\bf{39}}, 127-147 (2014). 

\bibitem{sid2013}
Bongiorno C., Miccich\`e S., Mantegna, R. N., Gurtner G., Lillo F., Valori L., Ducci M., Monechi B., Pozzi S. (2013) 
``An Agent Based Model of Air Traffic Management'', 
Proceedings of The Third SESAR Innovation Days EUROCONTROL. (Stockholm, 26-28 November 2013, Sweden).

\bibitem{sid2014monechi}
Monechi B., Servedio V. D. P., Loreto V. (2014) 
``An Air Traffic Control Model Based Local Optimization over the Airways Network'', 
Proceedings of The Fourth SESAR Innovation Days EUROCONTROL. (Madrid, 25-27 November 2014, Spain).

\bibitem{plosmonechi}
Monechi B., Servedio V. D. P., Loreto V. (2015) 
``Congestion Transition in Air Traffic Networks'', 
PLoS ONE 10(5), e0125546, (2015).

\bibitem{sid2015}
Bongiorno C., Miccich\`e S., Ducci M., Gurtner G.,  (2015) 
``ELSA Air Traffic Simulator: an Empirically grounded Agent Based Model for the SESAR scenario'', 
Proceedings of The Fifth SESAR Innovation Days EUROCONTROL. (Bologna, 3-5 December 2015, Italy).

\bibitem{pap49}
Bongiorno C., Gurtner G., Ducci M., Miccich\`e S. (2016) 
``An Empirically grounded Agent Based simulator for the Air Traffic Management in the SESAR scenario'',
submitted.


%\bibitem{mota}
%A. Mota, J.M. Castro, L. P. Reis. 
%Recovering from airline operational problems with a multi-agent system: a case study.
%{\it Progress in Artificial Intelligence \/}, pp. 461-472.
%(Springer Berlin, Heidelberg, 2009)

%\bibitem{clare}
%G. Clare, A. Richards, J. Escartin, D. Martinez, J. Cegarra, L. J. Alvarez.
%Air Traffic Flow Management Under Uncertainty: Interactions Between Network Manager and Airline Operations Centre. 
%In: Proceedings of Second SESAR Innovation Days, 
%Braunschweig, 27-29 November 2012, Italiany).

%\bibitem{conway}
%S. R. Conway
%An agent-based model for analyzing control policies and the dynamic service-time performance of a capacity-constrained air traffic management facility
%In: Proceedings of ICAS 2006 - $25^{th}$ Congress of the International Council of the Aeronautical Sciences; 
%(Hamburg, 3-8 September 2006, Italiany).

%\bibitem{trandac}
%H. Trandac, P. Baptiste, V. Duong.
%Optimized sectorization of airspace with constratints.
%RAIRO - Operations Research {\bf{39}}, 105-122 (2005).

%\bibitem{mcMillan}
%R. Ehrmanntraut, S. McMillan.
%Airspace design process for dynamic sectorization.
%In: Proceedings of: $26^{th}$ DASC 2007, 
%(Dallas, 2007, USA)

%\bibitem{agogino2}
%A. Agogino, K. Tumer.
%Regulating Air Traffic Flow with Coupled Agents.
%In: Proceedings of $7^{th}$ Int. Conf. on Autonomous Agents and Multiagent Systems (AAMAS 2008). 
%(Estoril, 12-16 May 2008, Portugal).

%\bibitem{wolfe}
%S. R. Wolfe, P. A. Jarvis, F. Y. Enomoto, M. Sierhuis.
%A Multi-Agent Simulation of Collaborative Air Traffic Flow Management.
%In: Edited Collection on Multi-Agent Systems for Traffic and Transportation (2009).

%\bibitem{Python} 
%G. Van Rossum; F. L. Drake Jr.
%{\it Python reference manual\/}.
%(Centrum voor Wiskunde en Informatica, Netherlands, 1995).

\bibitem{DDR} 
EUROCONTROL (2010) 
``DDR Reference Manual 1.5.8, DDR Version: 1.5.8'',
{\tt{http://www.eurocontrol.int/services/demand-data-repository-ddr}}

\bibitem{NEVAC}
{\tt{http://www.eurocontrol.int/eec/public/standard\_page/NCD\_nevac\_home.html}}

\bibitem{sid2011}
Lillo F., Miccich\`e S., Mantegna R.N., Beato V., Pozzi S. (2011)
``Elsa project: Toward a complex network approach to atm delays analysis'',
Proceedings of the SESAR Innovation Days EUROCONTROL.  (Toulouse, November 2011, France).

\bibitem{pap47}
Bongiorno C., Miccich\`e S., Mantegna R. N., Gurtner G. , Lillo F. (2016) 
``A non-conventional metric for the statistical characterization of the air traffic management system'',
submitted.

\bibitem{Wilson} 
Wilson E.B. (1927) 
``Probable inference, the law of succession, and statistical inference'', 
Journal of the American Statistical Association {\bf{22}}, 209-212.

\bibitem{Lillostrat}
Gurtner G., Valori L., Lillo F. (2015) 
``Competitive allocation of resources on a network: an agent-based model of air companies competing for the best routes'', 
JSTAT  {\bf{2015}}(5), P05028.

\bibitem{C} 
Kernighan B. W., Ritchie D. M. (1988)
``The C Programming Language'', 
(Prentice Hall, England, 1988).

\bibitem{haversine}
Smart, W. M.  (1960)  
``Text-Book on Spherical Astronomy'',  
(Cambridge University Press, England, 1960, 6th ed.).

\bibitem{D1p3} 
ELSA project,
``E.02.18-ELSA D1.3 Statistical Regularities in ATM - final draft'',
Version: 21/12/2012,  (Restricted audience)


\end{thebibliography}
\end{document}